\documentclass[ceurart]{ceurart}
\usepackage{comment}

\usepackage{graphicx}
\usepackage{amsmath}
\usepackage{listings}
\usepackage{caption}
\usepackage{subcaption}
\usepackage{hyperref}
\usepackage{multirow}
\usepackage{booktabs}
\usepackage{float}

\usepackage{graphicx}     % For including images
\usepackage{caption}      % For better control of figure captions
\usepackage{subcaption}   % For subfigures (if needed)
\usepackage{placeins}     % To prevent figures from floating into next sections (with \FloatBarrier)
\usepackage{subcaption}
\usepackage{graphicx}
\usepackage[justification=centering]{caption}

\definecolor{lightblue}{RGB}{173,216,230} % optional if not predefined

\begin{document}

\title{A Novel Hybrid Deep Learning Technique for Speech Emotion Detection using Feature Engineering}

 \author[1]{Shahana Yasmin Chowdhury}[
%  id=0000-0000-0000-0000,
 email=sychowdh@uno.edu]
  \author[2]{Bithi Banik}[
% orcid=0000-0000-0000-0000,
 email=bithi.banik@kristiania.no]
  \author[1]{Md Tamjidul Hoque}[
% orcid=0000-0000-0000-0000,
 email=thoque@uno.edu]
  \author[1]{Shreya Banerjee}[
% orcid=0000-0000-0000-0000,
 email=sbanerj1@uno.edu]
\address[1]{University of New Orleans, USA}
\address[2]{Kristiania University of Applied Sciences, Norway}

\begin{abstract}
Nowadays, \textit{speech emotion recognition} (SER) plays a vital role in the field of \textit{human-computer interaction} (HCI) and the evolution of \textit{artificial intelligence} (AI). Recent advancements in SER research have gained growing importance across various application areas, including healthcare, affective computing, personalized services, enhanced security, and AI-driven behavioral analysis. Existing machine learning (e.g., SVM, HMM) and deep learning (e.g., CNNs, LSTMs, Transformers) approaches have significantly advanced, proposing various algorithms for SER. However, these approaches often struggle to fully capture the sequential and contextual dependencies in speech signals, leading to suboptimal emotion classification, particularly in low-resource and noisy environments. To address this issue, we propose a novel sequence-based structured prediction framework that integrates Deep Conditional Random Fields (DeepCRF) with Bidirectional LSTM for detecting emotions in speech, which combines the benefits of deep feature learning with structured sequence prediction. Our proposed DCRF-BiLSTM model is used to recognize seven emotions: neutral, happy, sad, angry, fear, disgust, and surprise, which are trained on five datasets: RAVDESS (R), TESS (T), SAVEE (S), EmoDB (E), and Crema-D (C). The model achieves high accuracy on individual datasets, including 97.83\% on RAVDESS, 97.02\% on SAVEE, 95.10\% for CREMA-D, and a perfect 100\% on both TESS and EMO-DB. For the combined (R+T+S) datasets, it achieves 98.82\% accuracy, outperforming previously reported results. To our knowledge, no existing study has evaluated a single SER model across all five benchmark datasets (i.e., R+T+S+C+E) simultaneously. In our work, we introduce this comprehensive combination and achieve a remarkable overall accuracy of 93.76\%. These results confirm the robustness and generalizability of our DCRF-BiLSTM framework across diverse datasets.
\end{abstract}

% Todo: Pranish, refer to this
\conference{HHAI-WS 2025: Workshops at the Fourth International Conference on Hybrid Human-Artificial Intelligence (HHAI), June 9–13, 2025, Pisa, Italy}

\copyrightclause{Copyright for this paper by its authors. 
Use permitted under Creative Commons License Attribution 4.0 International (CC BY 4.0).}

\maketitle

\section{Introduction}

With technological advancements, research in \textit{human-computer interaction (HCI)} and \textit{artificial emotional intelligence (AEI)} is evolving rapidly \cite{krakovsky2018artificial, schuller2018age}. HCI \cite{hudlicka2003feel} focuses on how people communicate with computers and how effectively systems respond. Researchers aim to make interactions between humans and machines as natural as possible. Speech, being our primary form of communication, plays a key role in HCI, with microphones as auditory sensors \cite{mustaqeem2019cnn}. Through speech, we express emotions, making \textit{speech emotion recognition (SER)} essential for enhancing HCI \cite{chatterjee2018speech}.

\textit{Speech emotion detection (SED)} or SER is a sub-domain of \textit{natural language processing (NLP)} \cite{brown2020language, kusal2023systematic} and affective computing \cite{picard2000affective}. There has been a significant amount of published work in recent years \cite{madanian2023speech, akccay2020speech}. A conventional SER system uses various methods to extract and analyze speech signals to identify emotions \cite{akccay2020speech}. SER has many practical applications, especially in enhancing human-computer interactions by adding emotional awareness \cite{mustafa2018speech}.

One practical example of SER includes assessing call center agents' performance by detecting customer emotions such as anger or happiness. This feedback helps companies improve service quality, offer targeted training, and increase customer satisfaction and efficiency \cite{gadhe2015emotion}. Beyond this, many applications—including healthcare, smart home devices, audio surveillance, criminal investigations, recommendation systems, dialogue systems, and intelligent robots—benefit from detecting users' emotions through speech.

Despite advances, challenges in SER remain due to limited technology, the complex nature of emotions, and variations in language, accents, gender, and age, which all impact how emotions are expressed in speech \cite{radhika2025reliable, kowalczyk2012detecting}. To address these limitations, researchers have turned to cross-dataset integration techniques \cite{ottoni2023deep, milner2019cross}. By combining diverse datasets, models can learn more generalized patterns and improve recognition accuracy across multiple contexts \cite{jothimani2022mff}.

%In this research, we follow a multi-step process to enhance SER. First, we selected five publicly available and culturally diverse datasets: RAVDESS \cite{livingstone2018ryerson}, TESS \cite{SP2/E8H2MF_2020}, SAVEE \cite{jackson2014surrey}, EMO-DB \cite{burkhardt2005database}, and CREMA-D \cite{cao2014crema}. Next, we applied preprocessing steps to remove silence and resample the data. Data augmentation techniques were also used to increase robustness. Then, a feature selection process was conducted to extract consistent and informative features across datasets. Finally, we propose the DCRF-BiLSTM model, which integrates Bidirectional LSTMs with a Conditional Random Field (CRF) layer to improve sequential emotion prediction.

To address the challenges discussed above, our contributions are outlined below:
\begin{itemize}
    \item We used five datasets together and applied data augmentation before model training. This approach helps mitigate the risk of overfitting and improves the generalizability of emotion recognition models.
    
    \item To the best of our knowledge, our proposed framework, DCRF-BiLSTM, outperforms other methods across most of the datasets in the literature.

    \item Our contribution lies in striking a balance between high feature diversity and maintaining computational efficiency, ensuring our model captures nuanced emotional characteristics without overfitting.
\end{itemize}

The structure of this paper is as follows: Section 2 reviews existing literature on SER. 
%to understand current trends, gain insights, and identify areas for improvement. 
Section 3 highlights the importance of SED or SER. Section 4 provides an overview of our system, including discussions on datasets, data preprocessing, data augmentation techniques, feature extraction, feature selection, and our proposed architectures. In Section 5, we present a detailed and a comparative analysis of the experimental results using five datasets. Finally, Section 6 concludes with a discussion on current limitations and future research directions in SED.

\section{Related Work}

%%%%%%%%%%%%%%%%%%%%%%%%%%%%%%%%%%%%%%%%
%1st Paragraph: general discussion about SER, explain what SER is and its significance in HCI, mental health, virtual assistants with real world application.
SER identifies speakers' emotional states from speech and plays a vital role in enhancing HCI across domains like call centers, mental health diagnostics, and driver monitoring systems. By interpreting emotional cues, SER improves user interaction, enables emotionally adaptive interfaces, and supports stress or anxiety detection \cite{nantasri2020light, mukhamediya2023effect}.

%2nd paragraph: Traditional approach highlighting feature related work in literature. 
The traditional approach to SER has primarily emphasized extracting prosodic features—such as pitch, intensity, and energy—and spectral features like Mel-frequency cepstral coefficients (MFCC) and Linear Predictive Coefficients (LPC), which are essential for representing emotional cues in speech \cite{madanian2023speech}. These features have typically been evaluated using machine learning classifiers such as Gaussian Mixture Models and Support Vector Machines (SVM) \cite{er2020novel, mukhamediya2023effect}. While earlier studies focused on a limited set of handcrafted features for model development \cite{mukhamediya2023effect}, there remains limited research exploring a broader range of acoustic feature types. In contrast, our work leverages a more comprehensive and diverse set of features—including MFCC, Chroma, Spectral Contrast, RMSE, ZCR, and Log-Mel Spectrograms—resulting in a high-dimensional feature space. This extensive variation not only captures more nuanced emotional characteristics but may also contribute to improved classification performance compared to prior works that relied on fewer features.

%3rd paragraph: discuss Deep learning methods use in literature. CNNs, RNNs, LSTM/BiLSTM, attention mechanisms. Need to mention, how deep learning has improved SER but still leaves room for innovation.

Deep learning has revolutionized various fields, including SER, by providing advanced methods for processing and analyzing complex data. Techniques such as Convolutional Neural Networks (CNNs), Recurrent Neural Networks (RNNs), Long Short-Term Memory networks (LSTMs), and attention mechanisms have significantly enhanced the accuracy and efficiency of SER systems.  In SER, CNNs have been used to extract high-level features from spectrograms, which are then used for emotion classification \cite{mukhamediya2023effect}. RNNs are designed to process sequential data, making them suitable for time-series analysis, such as speech signals \cite{akccay2020speech}. LSTMs, a type of RNN, address the vanishing gradient problem and are capable of learning long-term dependencies, which is beneficial for SER \cite{jothimani2022mff}. Combining CNNs with LSTMs allows for the extraction of both spatial and temporal features, enhancing the performance of SER systems \cite{gupta2022detecting, akccay2020speech}. Deep learning has significantly improved the accuracy and robustness of SER systems by enabling the automatic extraction and processing of complex features from raw data \cite{mukhamediya2023effect}. 

%4th paragraph : discuss about the datasets used in the previous work.

The study of emotion recognition using speech datasets is a key area in AI and machine learning. Datasets like RAVDESS, TESS, SAVEE, EMO-DB, and CREMA-D provide diverse audio samples for training and evaluation. RAVDESS offers high-quality recordings and wide emotion coverage, achieving 97.16\% accuracy with ConvLSTM \cite{ben2024enhancing}, though its limited size may cause overfitting \cite{issa2020speech}. TESS has 2,800 samples and shows perfect accuracy in some studies \cite{ben2024enhancing}, but lacks speaker diversity \cite{ahmed2023ensemble}. SAVEE, focused on male British speech, reached 97.45\% accuracy \cite{ahmed2023ensemble, ben2024enhancing}, yet its small size is a limitation \cite{issa2020speech, ben2024enhancing}. EMO-DB supports cross-linguistic studies \cite{ben2024enhancing} with 535 German samples but faces overfitting risks \cite{issa2020speech, ben2024enhancing}. CREMA-D includes varied demographics \cite{ ahmed2023ensemble}, though its complexity impacts consistent training despite achieving 83.28\% in some works \cite{ben2024enhancing}. While valuable, these datasets highlight the need for more diverse and extensive data to improve SER model generalization.

%5th paragraph: summarize the gap ( due to feature selection criteria, model selection criteria,). Which lead to our proposed contribution. Conclude with how your model aims to fill these gaps.

Despite notable progress in deep learning-based SER, key challenges persist in achieving high accuracy and robust generalization across diverse datasets. Variations in emotional expression across individuals, languages, and cultural backgrounds make it difficult for models to generalize effectively \cite{ben2024enhancing}. The inherent subjectivity of emotions also introduces ambiguity in labeling, leading to inconsistencies in classification outcomes \cite{er2020novel}. Moreover, many existing SER datasets are scripted and limited in speaker and emotional diversity, which reduces their applicability to real-world scenarios and increases the risk of overfitting \cite{ahmed2023ensemble}. Issues such as class imbalance and small dataset sizes further hinder deep learning model performance and stability \cite{ahmed2023ensemble}.
While deep models like CNNs and LSTMs can automatically learn features, selecting relevant and informative features remains a challenge \cite{er2020novel}. Additionally, the computational demands of these architectures can limit their use in real-time systems. Although techniques like data augmentation and cross-dataset integration offer potential improvements by enhancing training diversity \cite{ben2024enhancing, ahmed2023ensemble}, relatively few studies explore the impact of high-dimensional and varied acoustic features in conjunction with structured prediction models. These limitations motivate our work, which addresses the need for better feature diversity, improved generalization, and context-aware modeling in SER.

%%%%%%%%%%%%%%%%%%%%%%%%%%%%%%%%%%%%%%%%

\section{Methodology}
The proposed framework begins with the preparation of datasets. After collecting five publicly available and widely used speech emotion datasets, we applied two key preprocessing steps: data preprocessing and data augmentation. Once the data was prepared, we proceeded with feature extraction and feature selection to identify the most informative attributes for emotion recognition. Following this, we detail the architecture of our proposed model. This section also outlines the processes involved in model training and evaluation

\begin{figure*}[hbt]
  \centering
  \includegraphics[width=0.8\linewidth]
{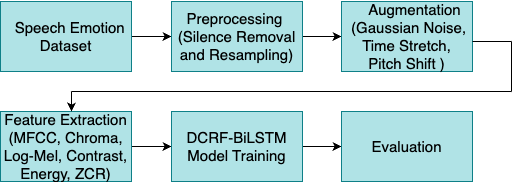}  %\includegraphics[width=0.8\linewidth]
  \caption{System overview of the proposed framework.}
  \label{fig:sysoverview}
\end{figure*}

Our proposed framework DCRF-BiLSTM is shown in 
%\textbf{\textit{\textcolor{blue}{
Figure~\ref{fig:sysoverview}
%}}}
. At first, we fetch all the audio or speech signals from the datasets, then we preprocess each audio signal with silence removal and resampling, followed by three types of Augmentation techniques, i.e, Gaussian noise, time stretch, pitch shift. At feature extraction steps, we extracted 6 different types of features with variation, and a total of 190 features were selected. For the model training, we applied a Deep Conditional Random Fields (DeepCRF) and Bidirectional LSTM framework to detect emotions in speech, combining deep feature learning with structured sequence prediction. Our proposed model is designed to recognize seven emotions — neutral, happy, sad, angry, fear, disgust, and surprise — and is trained on five datasets: RAVDESS, TESS, SAVEE, EmoDB, and Crema-D.

\subsection{Dataset}
In this study, we used five publicly available datasets namely Ryerson Audio-Visual Database of Emotional Speech and Song (RAVDESS), Toronto Emotional Speech Set (TESS), Surrey Audio-Visual Expressed Emotion (SAVEE), EmoDB and  CREMA-D, and for speech emotion detection. %\textbf{\textit{\textcolor{blue}{
Table~\ref{tab:dataset-emo-class}
%}}}
shows the emotion count of each of the five datasets and combined datasets class-wise \cite{ahmed2023ensemble}. All datasets used in this study are publicly available and widely used for academic research. We did not collect any new human data. Dataset diversity, spanning gender, age, and language, was considered to enhance fairness and generalizability.

%\begin{figure*}[hbt]
 % \centering
  %\includegraphics[width=0.45\linewidth]%{dcrf_bilstm/image/ravdess_emotions.png} \hfill
  %\includegraphics[width=0.45\linewidth]{dcrf_bilstm/image/tess_emotions.png} \hfill
  %\includegraphics[width=0.45\linewidth]{dcrf_bilstm/image/savee_emotions.png} \hfill
  %\includegraphics[width=0.45\linewidth]{dcrf_bilstm/image/emodb_emotions.png} \hfill
  %\includegraphics[width=0.45\linewidth]{dcrf_bilstm/image/cremaD_emotions.png} \hfill
  %\includegraphics[width=0.45\linewidth]{dcrf_bilstm/image/All_db_emotions.png}
  %\caption{Class-wise emotion count of five (RAVDESS, TESS, SAVEE, EmoDB, CREMA-D) datasets and the combined dataset.}
 % \label{fig:dataset-emo-class}
%\end{figure*}

\begin{table}[ht]
\caption{\label{dataset-distribution} Emotion class distribution across different datasets.}
\centering
\resizebox{\columnwidth}{!}{%
\begin{tabular}{lrrrrrrrr}
\hline
\rowcolor{lightgray}
\textbf{Dataset} & \textbf{Neutral} & \textbf{Happy} & \textbf{Sad} & \textbf{Angry} & \textbf{Fear} & \textbf{Disgust} & \textbf{Surprise} & \textbf{Total} \\
\hline
Ravdess   & 96  & 192 & 192 & 192 & 192 & 192 & 192 & 1248 \\ \hline
Tess      & 400 & 400 & 400 & 400 & 400 & 400 & 400 & 2800 \\  \hline
Savee     & 120 & 60  & 60  & 60  & 60  & 60  & 60  & 480  \\  \hline
EmoDB     & 79  & 71  & 62  & 127 & 69  & 46  & 0   & 454  \\  \hline
CremaD    & 1087& 1271& 1271& 1271& 1271& 1271& 0   & 7442 \\  \hline
R+T+S     & 616 & 652 & 652 & 652 & 652 & 652 & 652 & 4528 \\  \hline
R+T+S+E+C & 1782& 1994& 1985& 2050& 1992& 1969& 652 & 12424 \\  
\hline
\end{tabular}
}
 \label{tab:dataset-emo-class}
\end{table}

The description of the dataset are as follows:
    \begin{itemize}
        \item \textbf{Ryerson Audio-Visual Database of Emotional Speech and Song (RAVDESS}): This dataset\cite{livingstone2018ryerson} consists of 1440 audio wav files recorded by 12 males and 12 females. It includes emotions such as happy, calm, sad, fearful, angry, surprise, and disgust. The emotions are recorded at normal and strong intensity levels. It is a balanced dataset though the "neutral" class has fewer records compared to other classes. We exclude the calm emotion class from RAvDESS to matched with the combination dataset.
    \end{itemize}
    \begin{itemize}
\item \textbf{Toronto Emotional Speech Set (TESS)}: 
    A female-only dataset \cite{pichora2020toronto} consisting of 2800 audio files recorded by two actors aged 26 and 64 years. 
    The recordings portray seven emotions: happiness, anger, fear, disgust, pleasant surprise, sadness, and neutral.
\end{itemize}
\begin{itemize}
    \item \textbf{Surrey Audio-Visual Expressed Emotion (SAVEE)}: This dataset \cite{jackson2014surrey} consists of 120 audio clips recorded by four male speakers in the age group of 27 to 31 years. The emotions included are neutral, happy, sad, angry, disgust, fear and surprise. However, this dataset has a class imbalance issue, with the "neutral" class is almost double compared to all the other classes.
\end{itemize}
\begin{itemize}
    \item \textbf{EmoDB}: It \cite{burkhardt2005database} comprises 535 audio recordings in the German language categorized into seven emotional kinds: "anger," "fear," "sadness," "happiness," "disgust," "boredom," and "neutral".  The utterances are sampled at a rate of 16 kHz with a resolution of 16 bits. However, this dataset has a class imbalance issue, with the "anger," class utterance number is large compared to other classes.  We exclude the boredom emotion class from EmoDB to matched with the combination dataset.
\end{itemize}
\begin{itemize}
    \item \textbf{CREMAD}: The CREMA-D  \cite{cao2014crema} dataset contains 7,442 audio files recorded by 91 actors (48 males and 43 females) between the age group of 20 to 74 years. It consists of six emotions, namely happy, angry, neutral, sad, fear and disgust at four different emotions levels (low, medium, high, and unspecified)
\end{itemize}

\subsubsection{Data Preprocessing}
\begin{figure}[h!]
  \centering
  \begin{subfigure}[b]{0.44\linewidth}
    \includegraphics[width=\linewidth,height=4.8cm]{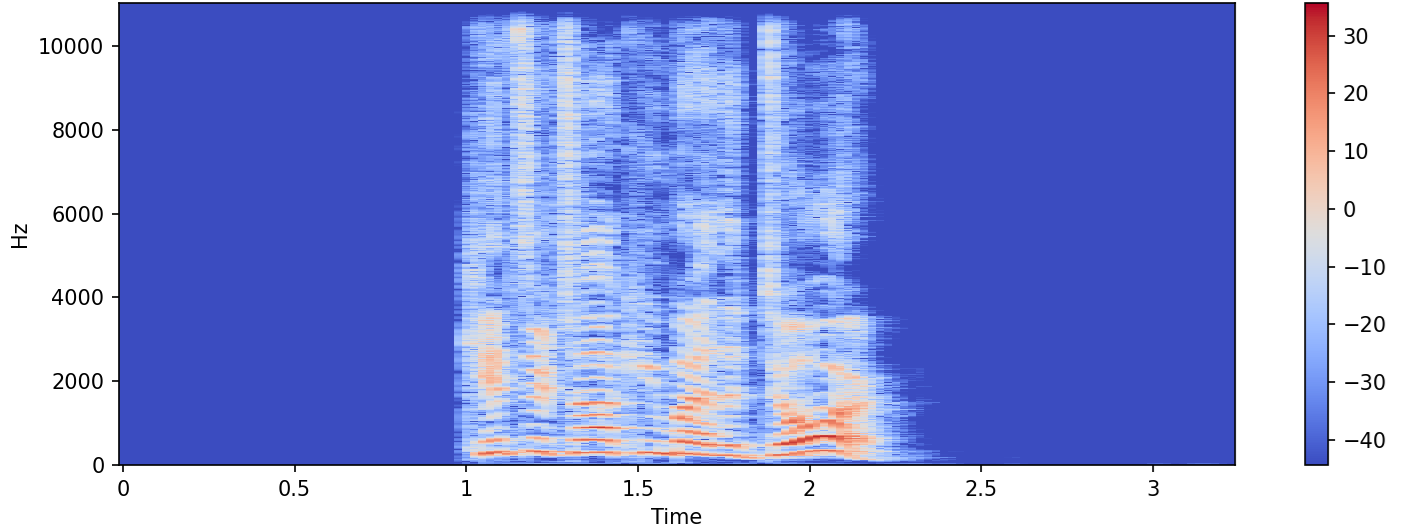}
    \caption{}
  \end{subfigure}
  \begin{subfigure}[b]{0.44\linewidth}
    \includegraphics[width=\linewidth,height=4.8cm]{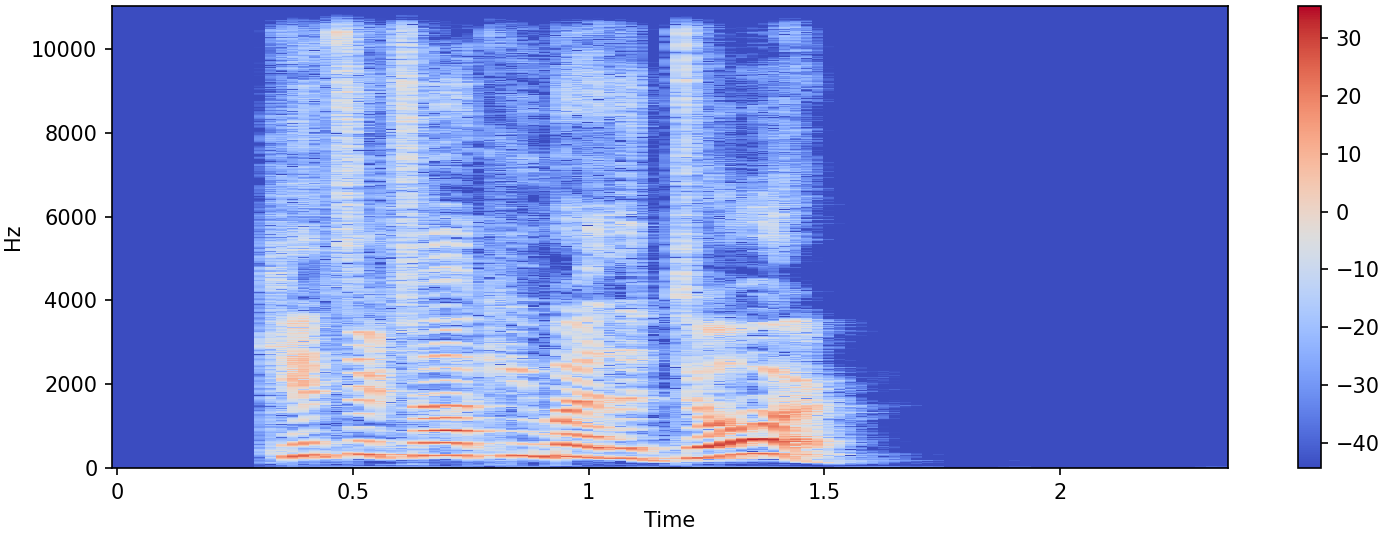}
    \caption{}
  \end{subfigure}
  %\caption [A Quantitative Versus Qualitative Problem]
  \caption{Spectrogram of \textit{happy} emotions (a) before silence removal, (b) after silence removal.
  }
  \label{fig:DPrp1}
\end{figure}

\begin{figure}[h!]
  \centering
  \begin{subfigure}[b]{0.465\linewidth}
    \includegraphics[width=\linewidth,height=4.8cm]{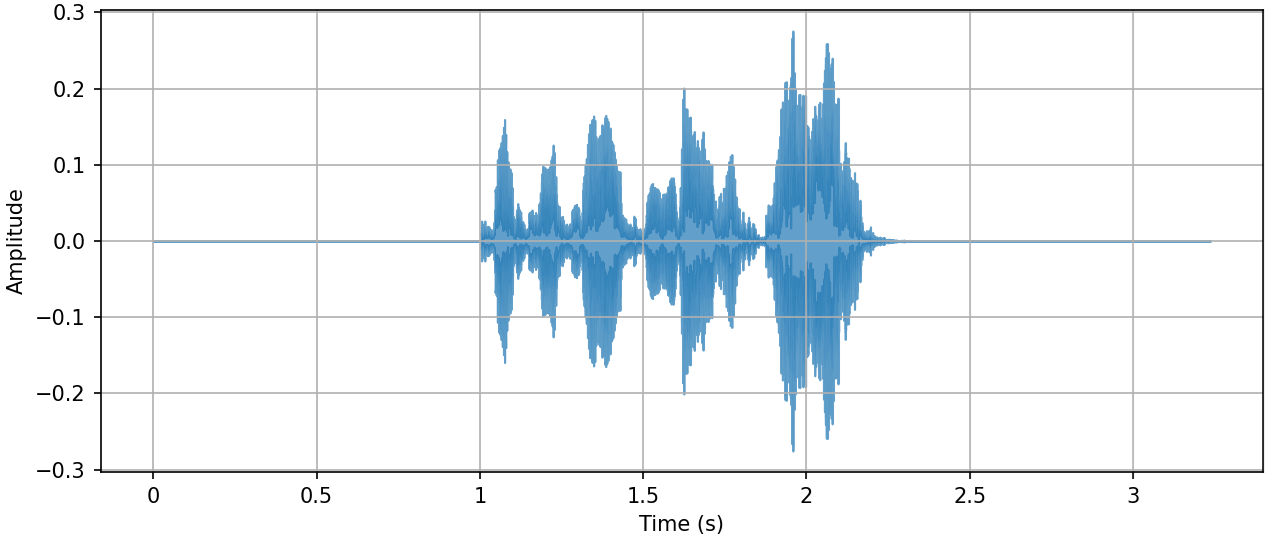}
    \caption{}
  \end{subfigure}
  \begin{subfigure}[b]{0.465\linewidth}
    \includegraphics[width=\linewidth,height=4.8cm]{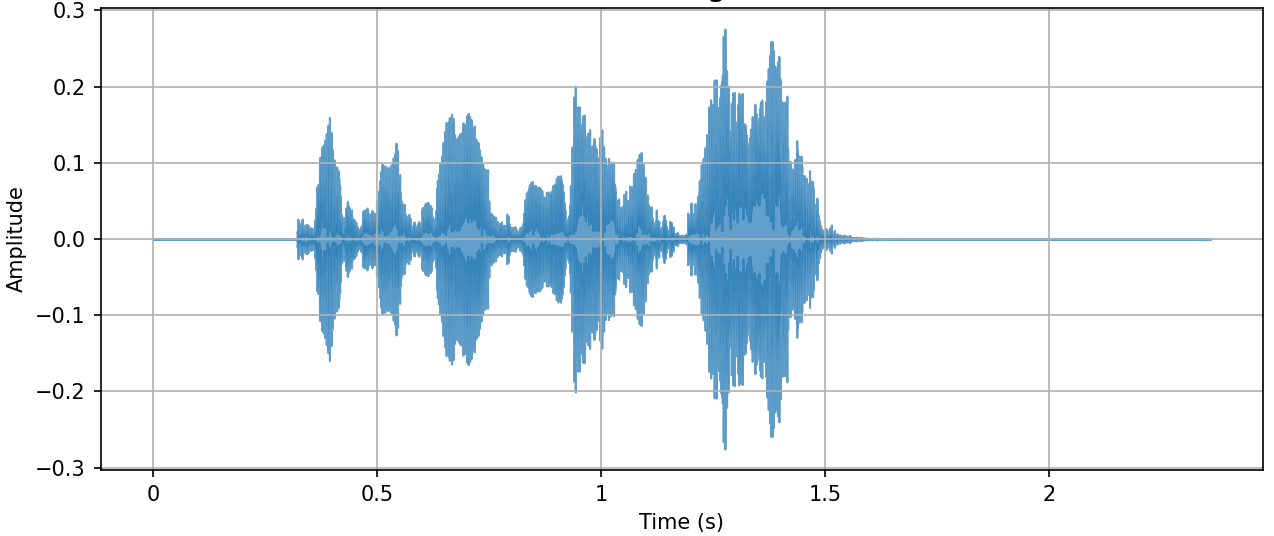}
    \caption{}
  \end{subfigure}
  \caption{Waveform of \textit{happy} emotions (a) before applying sampling rate, (b) after applying sampling rate.
  }
  \label{fig:DPrp2}
\end{figure}

Data Preprocessing is an important technique because it prepares the raw data for analysis and modeling. It also helps to extract relevant features, eliminate unnecessary information, and enhance model accuracy. As a part of data preprocessing, we removed unnecessary silence and resampled the data. Specifically, we removed 70\% of the silence from the beginning and end of each audio clip when the silence length is  more than 200ms. Additionally, we resampled each audio clip to a 22050 Hz sampling rate. Spectrogram and waveform of happy emotion with data preprocessing shown in 
%\textbf{\textit{\textcolor{blue}{
Figures~\ref{fig:DPrp1} - \ref{fig:DPrp2}
%}}}
, highlight the value of cleaning the data by removing silence, making the emotional characteristics of speech more prominent. It shows how preprocessing enhances the signal quality and prepares the data for more accurate emotion classification.

% \begin{figure*}[hbt]
%   \includegraphics[width=0.52\linewidth]{dcrf_bilstm/image/Original_happy_emotion_spectrogram.png} \hfill
%   \includegraphics[width=0.52\linewidth]{dcrf_bilstm/image/After_silenceRemove_resampling_spectrogram.png} 

%   \caption {Spectrogram of \textit{happy} emotions (a) before silence removal (b) after silence removal.}
%       \label{fig:DPrp1}
% \end{figure*}

% \begin{figure*}[hbt]

%   \includegraphics[width=0.52\linewidth]{dcrf_bilstm/image/Original_happy_emotion_waveform.png} \hfill
%     \includegraphics[width=0.52\linewidth]{dcrf_bilstm/image/After_silenceRemove_resampling.png}
%   \caption {Waveform of \textit{happy} emotions (a) before applying sampling rate (b) after applying sampling rate.}
%         \label{fig:DPrp2}
% \end{figure*}

\subsubsection{Data Augmentation}

\begin{figure*}[hbt]
  \centering  

  % --------- Row 1: 3 Images ----------
  \includegraphics[width=0.33\linewidth]{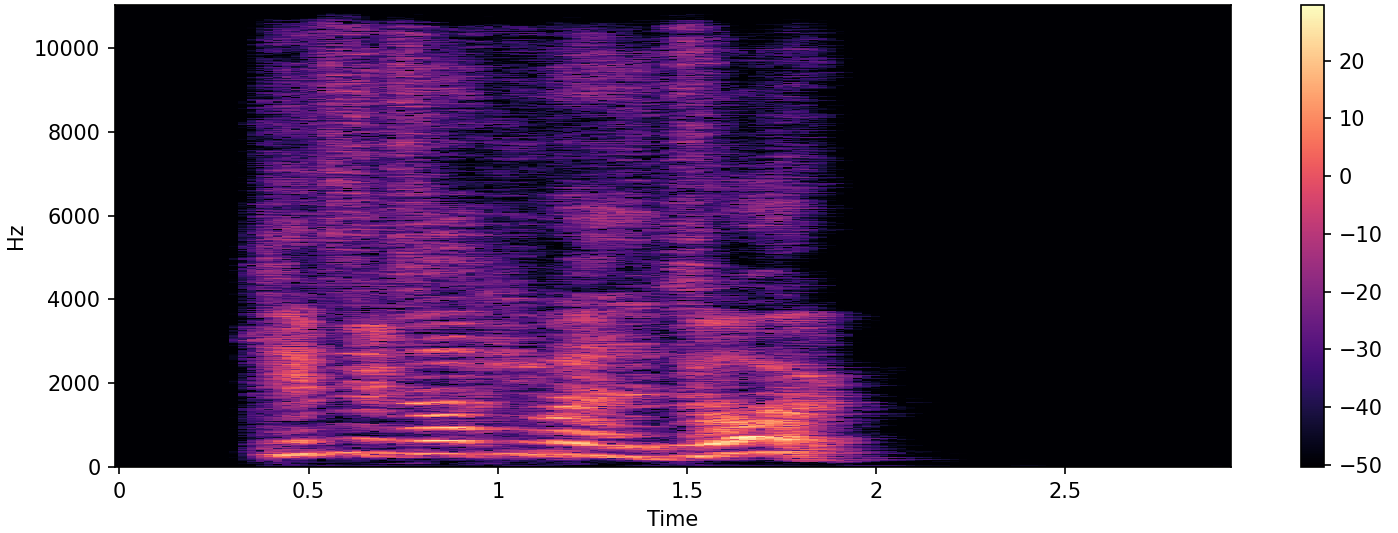} 
  \includegraphics[width=0.33\linewidth]{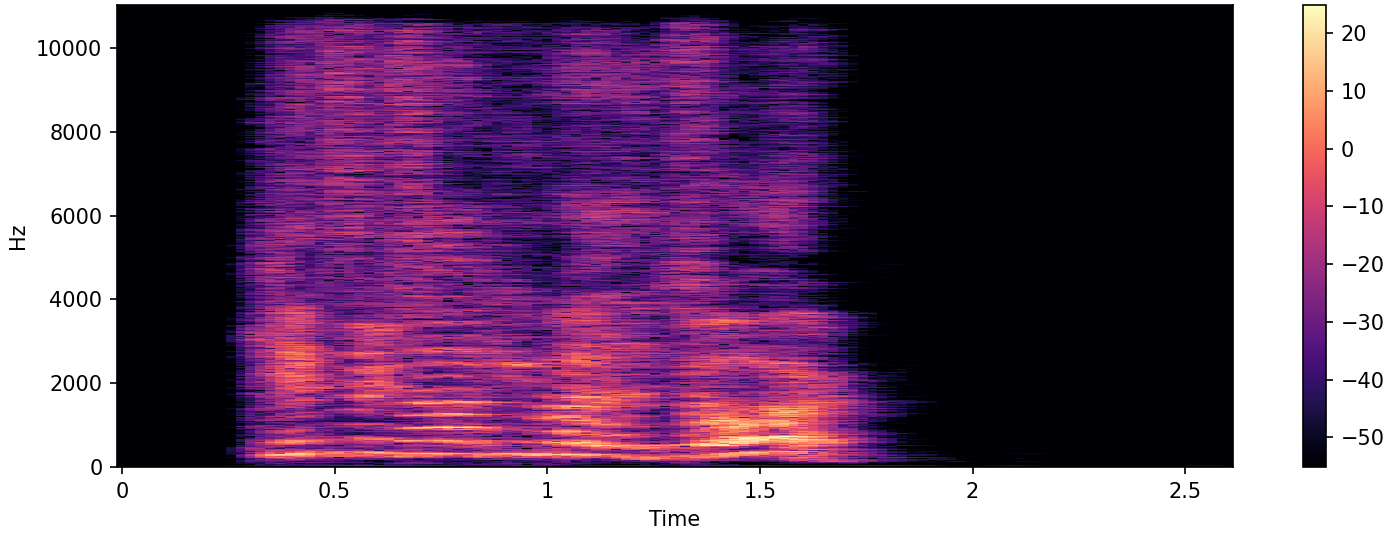} 
  \includegraphics[width=0.33\linewidth]{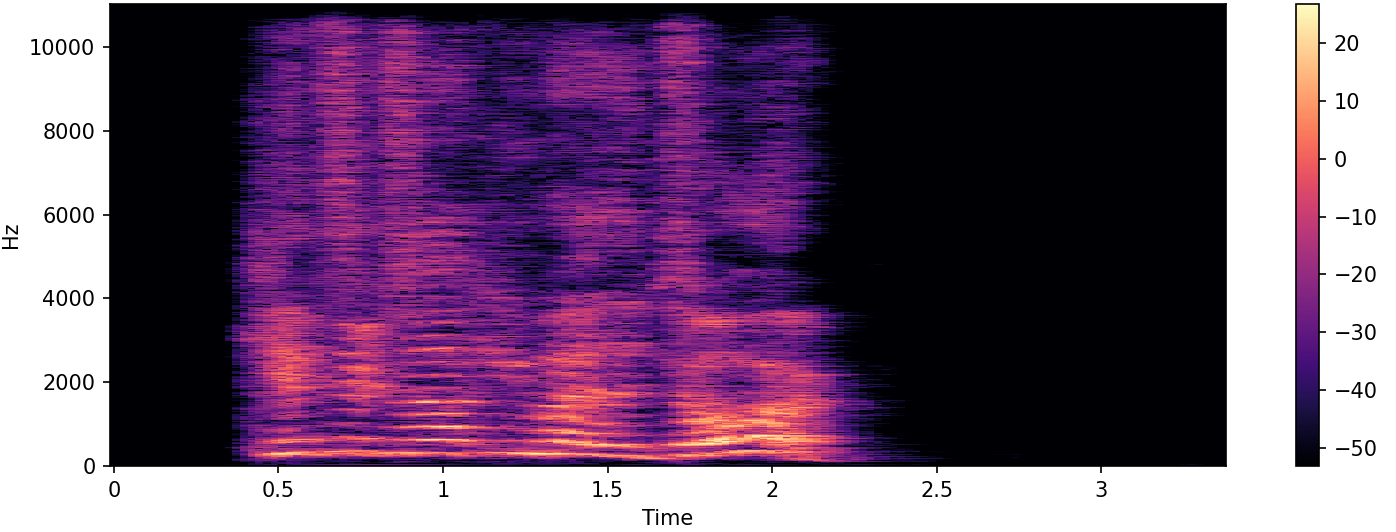} 

  \par\vspace{1em}  % Add space between the rows

  % --------- Row 2: 3 Images ----------
  \includegraphics[width=0.33\linewidth]{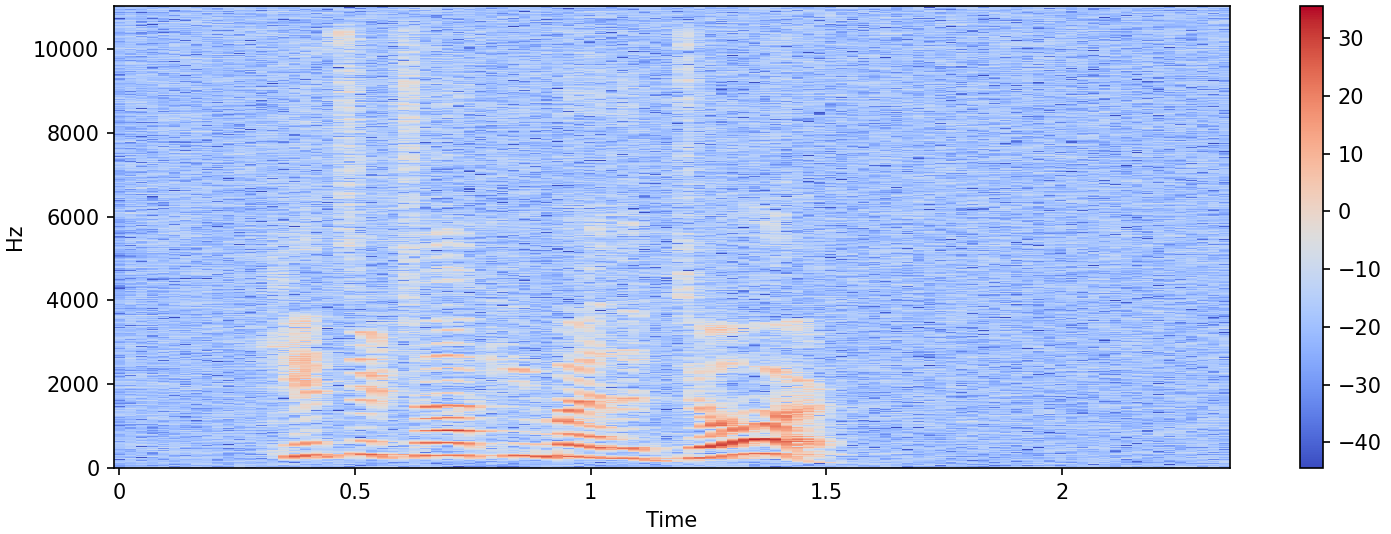} 
  \includegraphics[width=0.33\linewidth]{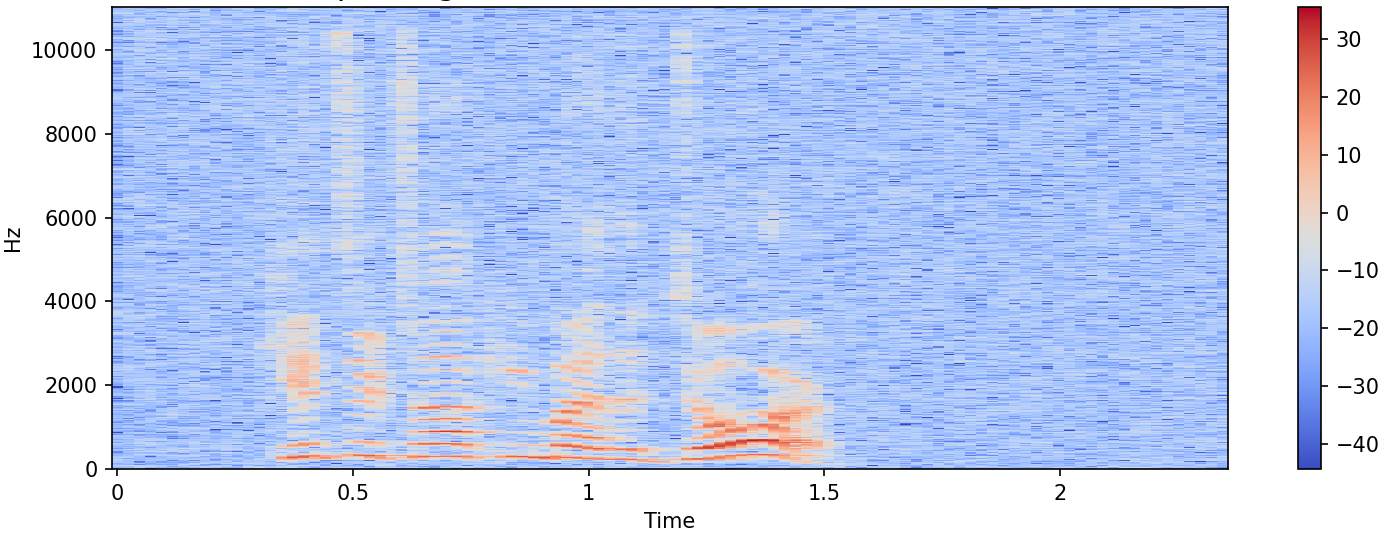} 
  \includegraphics[width=0.33\linewidth]{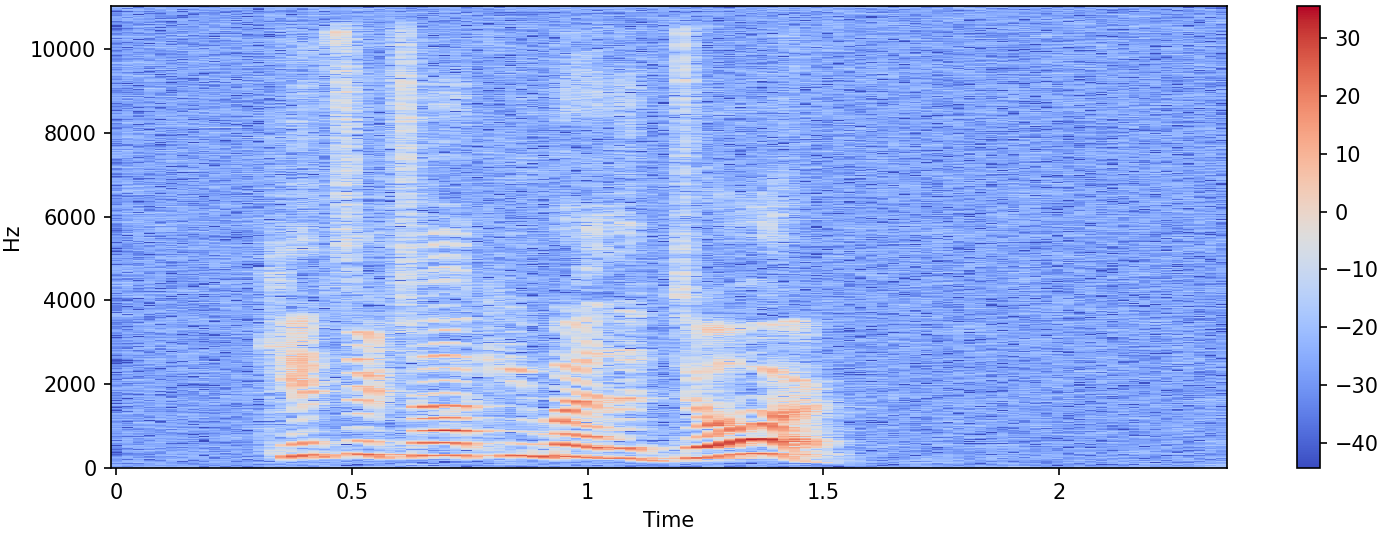}
  
  \par\vspace{1em}  % Add space between the rows

  % --------- Row 3: 3 Images ----------
  \includegraphics[width=0.33\linewidth]{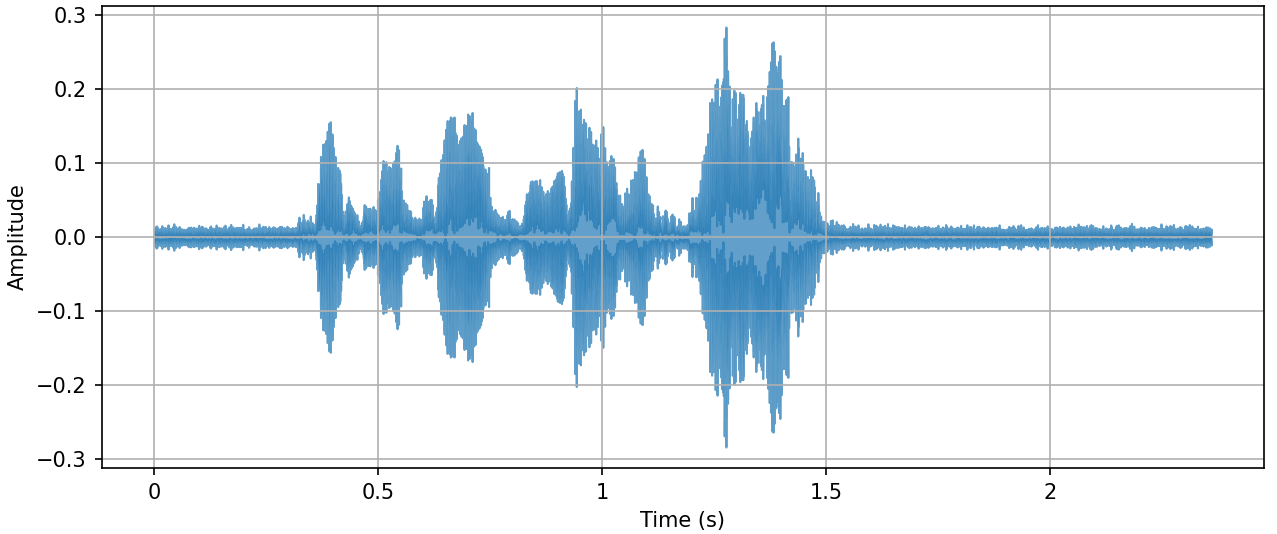} 
  \includegraphics[width=0.33\linewidth]{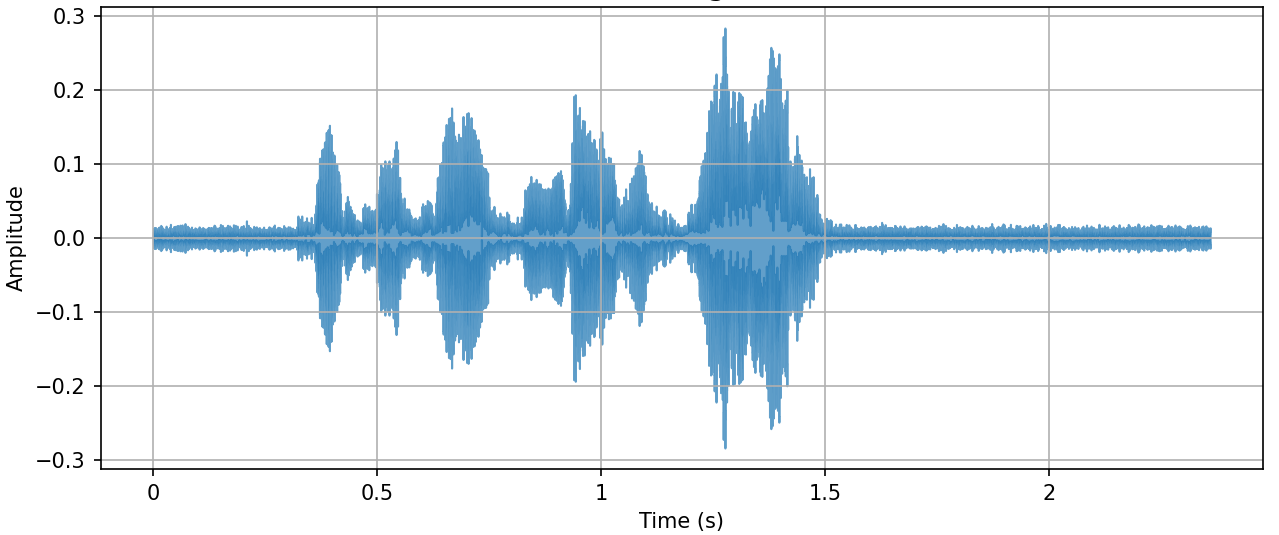} 
  \includegraphics[width=0.33\linewidth]{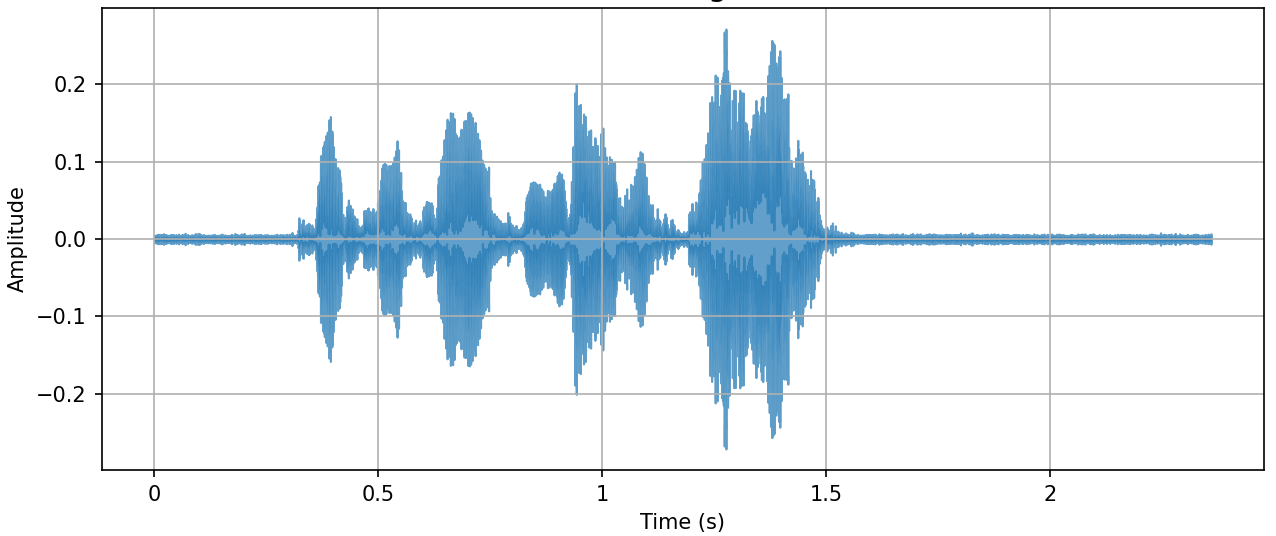} 

  \caption{Spectrograms and waveforms of augmented audio using time-stretching, pitch-shifting, and Gaussian noise—shown top-to-bottom, respectively, across three augmentation steps (left to right).}
  \label{fig:DA3}
\end{figure*}

To create variations in audio data, we applied the data augmentation process \cite{ferreira2022survey}. Data augmentation is a widely used technique in SER model to enhance the dataset, improve robustness and generalization. It’s also increased data size for proper training of the models. In this study, we noticed the data set imbalance over the emotion classes, so we injected Gaussian noise into  three faces. Our noise injection technique applied with the noise rate (0.035, 0.025, 0.015). We also applied two more augmentation techniques to our dataset, pitch shift and time stretch, by using librosa. The pitch shift technique shifts the pitch of an audio signal without changing its duration in three steps with the rate  parameter of (0.70, 0.80, 0.70). And we applied the time stretch technique to change the speed of an audio signal without altering its pitch with the speed factor rate (0.8, 0.9, 0.7). These techniques enhance the model's ability to handle different acoustic environments, improving the overall performance of the SER model. The impact of these augmentation techniques is visually shown in 
%\textbf{\textit{\textcolor{blue}{
Figure~\ref{fig:DA3}
%}}}
. Spectrogram and waveform visualizations illustrating the effects of different data augmentation techniques applied to a speech sample, Time-stretching increases the playback speed while preserving pitch, Pitch shifting alters the frequency content to simulate different speaker tones, 
and Gaussian noise injection simulates environmental noise to enhance model robustness. 

\subsection{Feature Extraction and Selection}
Feature extraction converted the raw data to a meaningful representation. It is an essential step in audio emotion detection, because raw data contains a vast amount of information. So, it helps to reduce the complexity of the data by capturing the most relevant information, which represents the emotional state in the speech. In this study, features have been extracted from WAV format audio files by exploiting Librosa, a Python package for music and audio analysis. In particular, we extracted and then selected the following 190 (= 80 + 36 + 64 + 6 + 3+ 1) features for our SER model. Selected features after extraction are shown in %\textbf{\textit{\textcolor{blue}{
Table~\ref{feature-list}
%}}}
.
\begin{itemize}
    \item \textbf{MFCCs (Mel-Frequency Cepstral Coefficients)} : The sounds produced by humans are shaped by the unique configuration of the vocal tract, including features such as the tongue and teeth. These structures influence each individual's distinct voice characteristics. Accurately capturing vocal tract shape is important, as it is reflected in the short-time power spectrum, typically represented by Mel-Frequency Cepstral Coefficients (MFCCs) \cite{gupta2022detecting}. MFCCs are widely used in SER research \cite{de2023ongoing, ahmed2023ensemble, singh2023speech}. To extract MFCCs, the speech signal is segmented into overlapping frames of 20–30 ms with a 10 ms shift to preserve temporal dynamics. Each frame is windowed and transformed using the Discrete Fourier Transform (DFT) to obtain magnitude spectrum \cite{nantasri2020light, hajarolasvadi20193d}. A set of 26 Mel-scaled filters is applied to mimic the human auditory system, producing 26 energy values per frame. These are converted into log filter bank energies. The relationship between Mel and physical frequency is expressed by %\textbf{\textit{\textcolor{blue}{
    Equation~\ref{eq:mfcc}
    %}}}
    .

\begin{equation} \label{eq:mfcc}
f_{\text{Mel}} = 2595 \cdot \log_{10} \left( 1 + \frac{f}{700} \right)
\end{equation}

Here, \( f \) denotes the physical frequency in Hertz (Hz), and \( f_{\text{Mel}} \) represents the perceived frequency by the human ear. Finally, the Discrete Cosine Transform (DCT) is applied to the log filter-bank energies to obtain the MFCCs.

For this study, 20 lower-dimensional MFCCs, 20 delta MFCCs, 20 delta-delta (delta\(^2\)) MFCCs, and 20 MFCC standard deviations were extracted from each audio file. The delta and delta-delta coefficients represent the first and second temporal derivatives of the MFCCs. These features effectively capture speech dynamics. MFCC envelopes are sufficient to reflect phoneme differences, enabling speech emotion recognition.

    \item \textbf{Chroma Features}: The Chroma feature captures the tonal content of audio by mapping frequency components to the 12 pitch classes of the chromatic scale \cite{birajdar2020speech}. It reflects energy distribution across pitch classes as a 12-dimensional vector. Chroma features are extracted using Short-Time Fourier Transform (STFT) on audio waveforms \cite{kumar2021cnn}. This study used three Chroma types: Chroma-STFT, Chroma-CQT (Constant-Q Transform), and Chroma-CENS (Chroma Energy Normalized Statistics). Each captures pitch information differently. From each audio file, 12 features were extracted per type, totaling 36 Chroma-related features. These represent tonal and harmonic speech content, aiding emotion recognition.
    \item \textbf{Log Mel Spectrogram (LMS)}: The Log Mel spectrogram improves audio spectrum representation by considering temporal changes and the human ear's frequency response. It includes signal segmentation, windowing, the Fourier transform, Mel filtering, and log compression to generate a logarithmic spectral view. The frequency spectrum is mapped onto Mel scale frequencies, forming a Mel spectrogram per window. Magnitude components of these frequencies were extracted using the Librosa library. The spectrogram visualizes signal intensity in the time-frequency domain using the Fast Fourier Transform (FFT) \cite{mukhamediya2023effect, meng2019speech}. It is essential for speech classification tasks and is effective with BiLSTM or DeepCRF. In this study, 64 LMS features were extracted from each audio file.

    \item \textbf{Spectral Contrast}: Spectral entropy serves as an indicator of voicing and signal quality \cite{kumar2021analysis}. It is often utilized to assess the 'speechiness' of an audio signal and is widely applied in distinguishing speech from background noise.
    \item \textbf{Root Mean Square Energy (RMSE)}: The Root Mean Square Energy gives the average signal amplitude over a frame, regardless of whether the values are positive or negative. It is useful for analyzing signal intensity in speech \cite{er2020novel}. In this study, RMSE values were extracted using the \texttt{Librosa} library. For a signal $x = x_1, x_2, ..., x_n$, the RMSE is calculated using %\textbf{\textit{\textcolor{blue}{
    Equation~\ref{eq:rmse}
    %}}}
    . In this study, we extracted 3 features for RMSE.

\begin{equation}
x_{RMS} = \sqrt{\frac{1}{n} \sum_{i=1}^{n} x_i^2}
\label{eq:rmse}
\end{equation}
\end{itemize}

\begin{table*}[hbt]
  \centering
  \caption{\label{feature-list}Extracted and selected feature List.}
  \begin{tabular}{lll}
    \hline
    \textbf{Feature} & \textbf{Description}  & \textbf{Number of Features}  \\
    \hline
    MFCCs	                & Temporal features of individual speech          & 20 \\
    MFCCs Delta	        & First derivative of MFCCs                        & 20 \\
    MFCCs Delta2	        & Second derivative of MFCCs                       & 20 \\
    MFCCstd	            & MFCCs standard deviation                         & 20 \\
    Chroma STFT, CQT, CENS & Tonal and harmonic structure analysis            & 36 \\
    Log Mel Spectrogram   & Captures both frequency and energy variations    & 64 \\
    Spectral Contrast     & Measures the difference in peak and valley energy& 6 \\
    Energy	            & Audio intensity of the signal                   & 3  \\
    ZCR	                & Measures signal sign changes over zero          & 1  \\
    \hline
  \end{tabular}
\end{table*}

\begin{itemize}
    \item \textbf{Zero Crossing Rate (ZCR)}: Zero Crossing Rate is a commonly used feature in SER. It counts the number of times a signal crosses the zero value in a given time frame. This helps to differentiate between voiced and unvoiced parts of speech \cite{ahmed2023ensemble}. In this work, we used the \texttt{Librosa} library to extract ZCR values from the audio datasets. The ZCR is defined mathematically in %\textbf{\textit{\textcolor{blue}{
    Equation~\ref{eq:zcr}
%    }}}
    %, where $s$ represents a signal of length $T$, and $1_{\mathbb{R}<0}$ is an indicator function that returns 1 if its argument is less than 0, otherwise 0.

    \begin{equation} \label{eq:zcr}
ZCR = \frac{1}{T-1} \sum_{t=1}^{T-1} 1_{\mathbb{R}<0} \left( s_t \cdot s_{t+1} \right)
\end{equation}

\end{itemize}

\subsection{Proposed Architecture}

In this study, we utilize the CRF (Conditional Random Field) layer, a specialized layer used for sequence tagging and prediction tasks, particularly in NLP, which exists within the TensorFlow Addons (TFA) module. We combined the DeepCRF with Bi-LSTM to train our model. %\textbf{\textit{\textcolor{blue}{
Table~\ref{tab:model_summary}
%}}}
shows our model architecture.

The architecture of the model is created with a sequential layer, which builds a Bidirectional LSTM layers to learn hierarchical and contextualized temporal features in the audio or speech data from both forward and backward directions. It uses 512 LSTM units, with L2 regularization to reduce overfitting, and Batch Normalization to accelerate training. A dropout rate of 0.3 is applied to prevent overfitting. In the second and third layer, same Bi-LSTM with 512 units is added for deeper feature extraction. Similar Batch Normalization and Dropout layers are applied.

The next two layers are a Dense layer with 512 units and Swish activation, which improves performance by addressing the vanishing gradient problem. LeakyReLU is applied at the end for additional non-linearity.  After that, we include the DeepCRF Layer with LeakyReLU activation to model the sequential structure of speech data. CRF performs best for sequence labeling tasks because it considers the context of neighboring predictions.

The final layer is the output layer. We use a TimeDistributed layer that applies a Dense layer with the number of classes and a Softmax activation to generate class probabilities for each time step. A reshape layer is also used to ensure the output matches the shape of the class number by representing the predicted emotion. This model is designed for speech emotion recognition by extracting high-level features using LSTMs and leveraging the CRF layer to enhance prediction accuracy through temporal relationships in speech signals.

%\begin{figure}[hbt]
%  \centering
%  \begin{minipage}%{0.9\columnwidth}
%    \centering
    %\includegraphics[width=\linewidth]{dcrf_bilstm/image/summary_bold.png}
%    \caption{Layer-wise architecture of the proposed DCRF-BiLSTM model. This summary outlines the input and output dimensions at each layer, showing the stacked Bidirectional LSTM layers followed by Batch Normalization, Dropout, Dense, and CRF layers. The final TimeDistributed layer outputs class probabilities for each time step. This architecture is crucial for capturing temporal patterns and improving sequence-based emotion classification.}
%    \label{fig:modelsummary}
%  \end{minipage}
% \end{figure}

\begin{table}[h!]
\centering
\caption{Layer-wise architecture of the proposed DCRF-BiLSTM model. This summary outlines the input and output dimensions at each layer, showing the stacked Bidirectional LSTM layers followed by Batch Normalization, Dropout, Dense, and CRF layers. The final TimeDistributed layer outputs class probabilities for each time step. This architecture is crucial for capturing temporal patterns and improving sequence-based emotion classification.}
\label{tab:model_summary}
\begin{tabular}{|l|c|c|}
\hline
\rowcolor{LightYellow}
\multicolumn{3}{|l|}{\textbf{Model: "DCRF-BiLSTM"}} \\ \hline
\textbf{Layer (Type)} & \textbf{Output Shape} & \textbf{Param \#} \\ \hline
bidirectional (Bidirectional) & (None, 1, 1024) & 2879488 \\ \hline
batch\_normalization (BatchNormalization) & (None, 1, 1024) & 4096 \\ \hline
dropout (Dropout) & (None, 1, 1024) & 0 \\ \hline
bidirectional\_1 (Bidirectional) & (None, 1, 1024) & 6295552 \\ \hline
batch\_normalization\_1 (BatchNormalization) & (None, 1, 1024) & 4096 \\ \hline
dropout 1 (Dropout) & (None, 1, 1024) & 0 \\ \hline
bidirectional\_2 (Bidirectional) & (None, 1, 1024) & 6295552 \\ \hline
batch\_normalization\_2 (BatchNormalization) & (None, 1, 1024) & 4096 \\ \hline
dropout 2 (Dropout) & (None, 1, 1024) & 0 \\ \hline
dense (Dense) & (None, 1, 512) & 524800 \\ \hline
dropout 3 (Dropout) & (None, 1, 512) & 0 \\ \hline
dense 1 (Dense) & (None, 1, 512) & 262656 \\ \hline
dropout 4 (Dropout) & (None, 1, 512) & 0 \\ \hline
leaky\_re\_lu (LeakyReLU) & (None, 1, 512) & 0 \\ \hline
crf\_layer (CRFLayer) & (None, 1, 512) & 49 \\ \hline
time\_distributed (TimeDistributed) & (None, 1, 7) & 3591 \\ \hline \hline
\textbf{Total Parameters} & \multicolumn{2}{c|}{16273976 (62.08 MB)} \\ \hline
\textbf{Trainable Parameters} & \multicolumn{2}{c|}{16267832 (62.06 MB)} \\ \hline
\textbf{Non-trainable Parameters} & \multicolumn{2}{c|}{6144 (24.00 KB)} \\ \hline
\end{tabular}
\end{table}

\subsection{Model Training}
To ensure an effective evaluation and comparison with previous studies, the target datasets were divided into two parts every time. After preparing the feature set, we perform the data normalization and split the feature data into 80\% training data and 20\% testing data . The model was trained on the Augmented dataset using 500 epochs. To tune the hyperparameter, we used a learning rate of 0.0001, batch size 256, optimizer ‘Adam’ with loss function categorical- crossentropy. The whole process is developed using TensorFlow and Keras to build and train a deep learning model for our SER framework. Experiments were conducted using CPU(11th Gen Intel(R) Core(TM) i7-1165G7 @ 2.80GHz 2.80 GHz) and RAM (16GB); training each fold took approximately 7 hours.

\section{Experiments and Results}

% In this section, we present the results of our SED model and then analyze them. 
The experiment involved training the SED model on five individual datasets as well as the combined datasets (R+T+S) and (R+T+S+E+C) to assess its generalization abilities. The results were compared and analyzed using performance evaluation metrics such as accuracy, loss rate, and confusion matrix.

\subsection {Performance Matrices and Evaluation}

The performance of our model, DeepCRF, with the combination of Bi-LSTM is given in 
Table ~\ref{acc-result1}. Every time, each dataset was partitioned into train and test sets employing an 80\%-20\% split to assess the model’s performance on unseen data. The accuracy metric determines the percentage of samples predicted correctly by the model. Our proposed model has an accuracy of 100.00\% for TESS, 100.00\% for EMO-DB, 97.83\% for RAVDESS, 97.02\% for SAVEE, and 95.10\% for CREMA-D datasets. Furthermore, 98.82\% for combined RAVDESS, TESS, and SAVEE (R+T+S), 93.76\% for combined 5 datasets (R+T+S+E+C). 

%%Added by Bithi
We also performed 5-fold cross-validation with PCA-reduced features (principle component analysis) to evaluate model consistency \cite{abdi2010principal, benba2017voice}. As seen in Table~\ref{acc-result1}, the results remained highly stable across datasets, confirming the robustness of DeepCRF. The close agreement between 80-20 split and cross-validation accuracies highlights the model’s generalization ability, while PCA successfully reduced feature size without loss of accuracy.

\begin{table}[ht]
\centering
\caption{\label{acc-result1}Comparison of model accuracy using all features vs PCA-reduced features with 80-20 split and 5-fold cross-validation.}
\small % Reduce font size
\resizebox{\textwidth}{!}{%
\begin{tabular}{>{\raggedright\arraybackslash}p{2.2cm}|cc|cc|cc}
\hline
\rowcolor{gray!10}
\textbf{Datasets} 
& \multicolumn{2}{c|}{\cellcolor{cyan!8}\textbf{80-20 split}} 
& \multicolumn{2}{c|}{\cellcolor{green!10}\textbf{80-20 split using PCA}} 
& \multicolumn{2}{c}{\cellcolor{magenta!10}\textbf{5-fold cross-validation using PCA}} \\
\rowcolor{gray!10}
& Accuracy & Features & Accuracy & Reduced Features & Accuracy & Reduced Features  \\
\hline
Ravdess & 97.83\% & 190 & 97.88\% & 97 & 97.99\% & 97  \\ \hline
Tess & 100.00\% & 190 & 100.00\% & 97 & 99.99\% & 97  \\ \hline
Savee & 97.02\% & 190 & 97.17\% & 89 & 96.10\% & 89  \\ \hline
EmoDB & 100.00\% & 190 & 100.00\% & 82 & 99.26\% & 82  \\  \hline
CremaD & 95.10\% & 190 & 94.08\% & 95 & 94.41\% & 95  \\ \hline
\rowcolor{gray!10}
Combined (R+T+S) & 98.82\% & 190 & 99.15\% & 97 & 99.18\% & 97  \\  \hline
\rowcolor{gray!10}
Combined (R+T+S+E+C) & 93.76\% & 190 & 94.01\% & 97 & 94.03\% & 97  \\
\hline

\end{tabular}
}
\end{table}

%\begin{table}[ht]
%\centering
%\small
%\caption{\label{acc-result1}Results on proposed models based on the SER performance on RAVDESS, TESS, EMO-DB, SAVEE, and CREMA-D datasets.}
%\resizebox{0.3\textwidth}{!}{%
%\begin{tabular}{l|r}     \hline
    %\rowcolor{lightblue}
    %\textbf{Datasets} & \textbf{Accuracy} \\ \hline
    %Ravdess & 97.83\% \\ \hline
    %Tess & 100.00\% \\ \hline
    %Savee & 97.02\% \\ \hline
    %EmoDB & 100.00\% \\ \hline
    %CremaD & 95.10\% \\ \hline
    %Combined (R+T+S) & 98.82\% \\ \hline
    %Combined (R+T+S+E+C) & 93.76\% \\ \hline
%\end{tabular}
%} \end{table}

%\input{dcrf_bilstm/report/Ravdess_classification_report}
%\input{dcrf_bilstm/report/Tess_classification_report}
%\input{dcrf_bilstm/report/Savee_classification_report}
%\input{dcrf_bilstm/report/EmoDB_classification_report}
%\input{dcrf_bilstm/report/CremaD_classification_report}
\begin{table}[ht]
\centering
\caption{Class-wise emotion recognition performance comparison across RAVDESS, TESS, and SAVEE datasets.}
\label{tab:RTSinduv}

\small % Reduce font size
\resizebox{\textwidth}{!}{%
\begin{tabular}{>{\raggedright\arraybackslash}p{2.2cm}|ccc|ccc|ccc}
\hline
\rowcolor{gray!20}
\textbf{Emo/Metric} 
& \multicolumn{3}{c|}{\cellcolor{blue!10}\textbf{RAVDESS}} 
& \multicolumn{3}{c|}{\cellcolor{green!10}\textbf{TESS}} 
& \multicolumn{3}{c}{\cellcolor{orange!20}\textbf{SAVEE}} \\
\rowcolor{gray!10}
& Precision & Recall & F1-Score & Precision & Recall & F1-Score & Precision & Recall & F1-Score \\
\hline
Neutral         & 0.98 & 0.98 & 0.98 & 1.00 & 1.00 & 1.00 & 0.95 & 0.99 & 0.97 \\
Happy           & 0.99 & 0.96 & 0.98 & 1.00 & 1.00 & 1.00 & 0.99 & 0.96 & 0.98 \\
Sad             & 0.97 & 0.97 & 0.97 & 1.00 & 1.00 & 1.00 & 0.98 & 0.96 & 0.97 \\
Angry           & 0.98 & 0.99 & 0.99 & 1.00 & 1.00 & 1.00 & 1.00 & 0.99 & 0.99 \\
Fear            & 0.98 & 0.98 & 0.98 & 1.00 & 1.00 & 1.00 & 0.96 & 0.95 & 0.96 \\
Disgust         & 0.99 & 0.97 & 0.98 & 1.00 & 1.00 & 1.00 & 0.97 & 0.92 & 0.94 \\
Surprise        & 0.96 & 0.99 & 0.97 & 1.00 & 1.00 & 1.00 & 0.95 & 0.99 & 0.97 \\
\hline
Accuracy        &    &      & 0.98 &     &     & 1.00 &     &   & 0.97 \\
Macro Avg       & 0.98 & 0.98 & 0.98 & 1.00 & 1.00 & 1.00 & 0.97 & 0.97 & 0.97 \\
Weighted Avg    & 0.98 & 0.98 & 0.98 & 1.00 & 1.00 & 1.00 & 0.97 & 0.97 & 0.97 \\
\hline
\end{tabular}
}
\end{table}

\begin{table*}[ht]
\centering
\caption{Class-wise emotion recognition performance on EmoDB and CREMA-D datasets.}
\label{tab:emodb-cremad}
\small
\rowcolors{2}{white}{gray!10}
\begin{tabular}{l|ccc|ccc}
\hline
\rowcolor{lightgray}
\textbf{Emotion / Metric} &
\multicolumn{3}{c|}{\cellcolor{cyan!20} \textbf{EmoDB}} &
\multicolumn{3}{c}{\cellcolor{orange!25} \textbf{CREMA-D}} \\
\rowcolor{lightgray}
& Precision & Recall & F1-score & Precision & Recall & F1-score \\
\hline
Neutral      & 1.00 & 1.00 & 1.00 & 0.94 & 0.96 & 0.95 \\
Happy        & 1.00 & 1.00 & 1.00 & 0.97 & 0.94 & 0.96 \\
Sad          & 1.00 & 1.00 & 1.00 & 0.92 & 0.96 & 0.94 \\
Angry        & 1.00 & 1.00 & 1.00 & 0.97 & 0.98 & 0.97 \\
Fear         & 1.00 & 1.00 & 1.00 & 0.96 & 0.93 & 0.94 \\
Disgust      & 1.00 & 1.00 & 1.00 & 0.95 & 0.93 & 0.94 \\
\hline
Accuracy     &   &  & 1.00 &     &      & 0.95 \\
Macro Avg    & 1.00 & 1.00 & 1.00 & 0.95 & 0.95 & 0.95 \\
Weighted Avg & 1.00 & 1.00 & 1.00 & 0.95 & 0.95 & 0.95 \\
\hline
\end{tabular}
\end{table*}

\begin{table*}[ht]
\centering
\caption{Class-wise emotion recognition performance on Combined R+T+S and R+T+S+E+C datasets.}
\label{tab:rts-rtsec}
\small
\rowcolors{2}{white}{gray!10}
\begin{tabular}{l|ccc|ccc}
\hline
\rowcolor{lightgray}
\textbf{Emotion / Metric} &
\multicolumn{3}{c|}{\cellcolor{blue!15} \textbf{Combined R+T+S}} &
\multicolumn{3}{c}{\cellcolor{green!20} \textbf{Combined R+T+S+E+C}} \\
\rowcolor{lightgray}
& Precision & Recall & F1-score & Precision & Recall & F1-score \\
\hline
Neutral      & 0.99 & 0.99 & 0.99 & 0.92 & 0.95 & 0.93 \\
Happy        & 0.98 & 0.98 & 0.98 & 0.94 & 0.93 & 0.94 \\
Sad          & 0.99 & 0.99 & 0.99 & 0.91 & 0.94 & 0.92 \\
Angry        & 0.99 & 0.99 & 0.99 & 0.95 & 0.98 & 0.96 \\
Fear         & 0.99 & 0.98 & 0.99 & 0.95 & 0.90 & 0.92 \\
Disgust      & 0.99 & 1.00 & 0.99 & 0.94 & 0.92 & 0.93 \\
Surprise     & 0.99 & 0.98 & 0.99 & 0.98 & 0.99 & 0.98 \\
\hline
Accuracy     &   &   & 0.99 &      &       & 0.94 \\
Macro Avg    & 0.99 & 0.99 & 0.99 & 0.94 & 0.94 & 0.94 \\
Weighted Avg & 0.99 & 0.99 & 0.99 & 0.94 & 0.94 & 0.94 \\
\hline
\end{tabular}
\end{table*}

We analyze the performance of five individual datasets and the combined dataset using weighted average (WA) and macro average (MA), due to class imbalance. We also report the precision, recall, and F1-score of the proposed model on individual datasets for the same reason. Tables \ref{tab:RTSinduv}-\ref{tab:rts-rtsec} show class-wise emotion recognition performance for individual and combined datasets.

\begin{figure*}[hbt]
  \centering
  \includegraphics[width=0.65\linewidth]{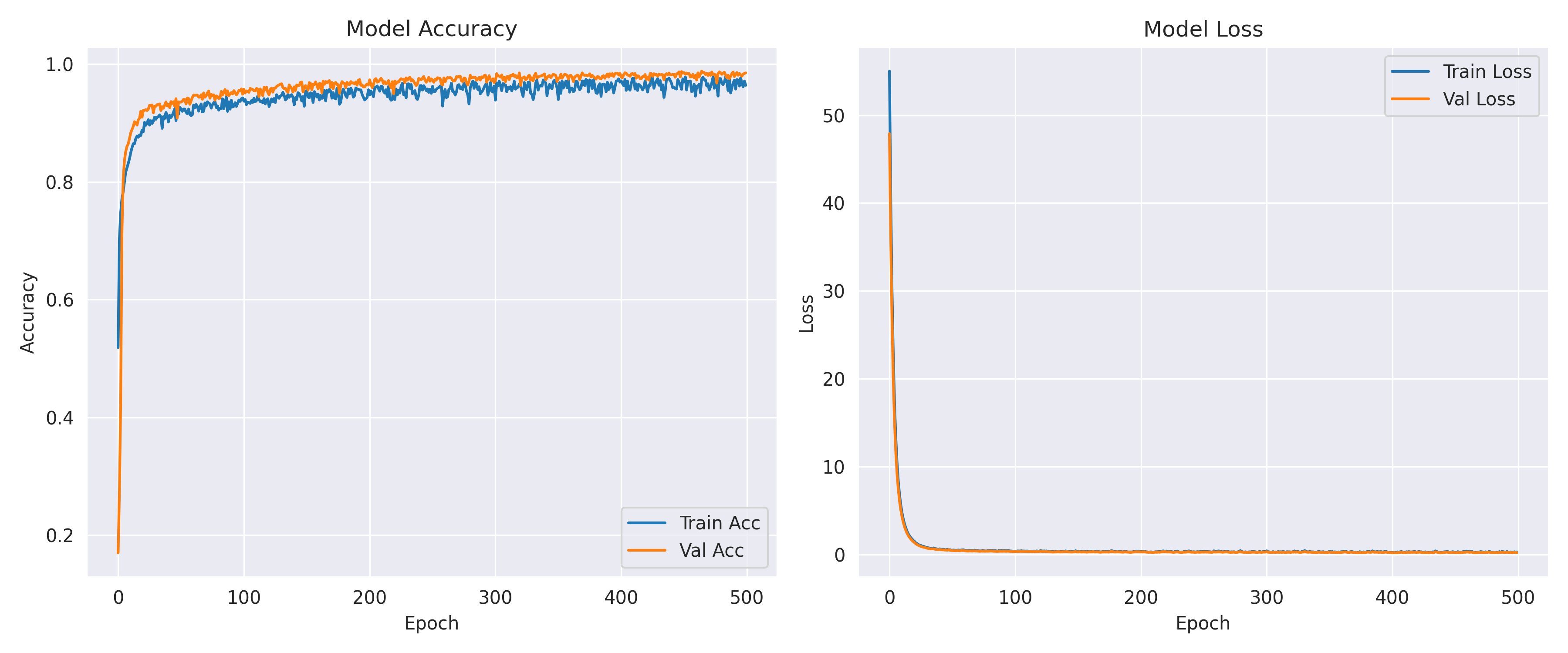}
  %\vspace{0.5cm}
  \includegraphics[width=0.65\linewidth]{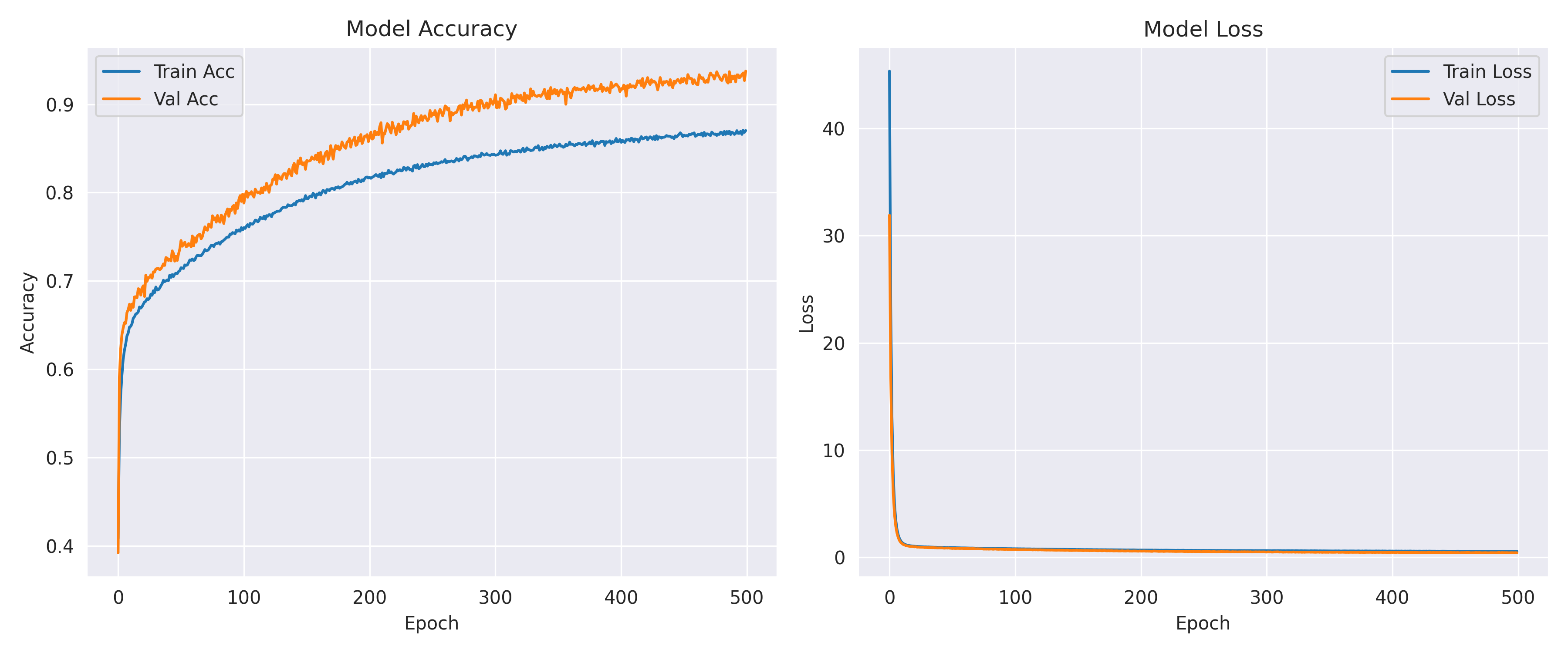}
  \caption{Training vs. validation accuracy and training loss vs validation loss curves for the DCRF-BiLSTM model. Top: Combined (R+T+S). Bottom: Combined (R+T+S+E+C).}
  \label{fig:accuracy_curves}
\end{figure*}

High accuracy in recognizing speech emotion is the model's primary goal, so this work records only the best-fit model. Validation accuracy is the key indicator of model generalization. When validation accuracy reaches its peak, prediction is optimal with this model. Figure~\ref{fig:accuracy_curves} shows results predicted by the proposed model for the two combined datasets. Appendix A : Figure~\ref{fig:ravdess_curve}-\ref{fig:cremad_curve} includes validation accuracy and loss curves for the five individual datasets.

\begin{figure*}[hbt]
  \centering  

  % --------- Row 1: 4 Images ----------
  \includegraphics[width=0.25\linewidth]{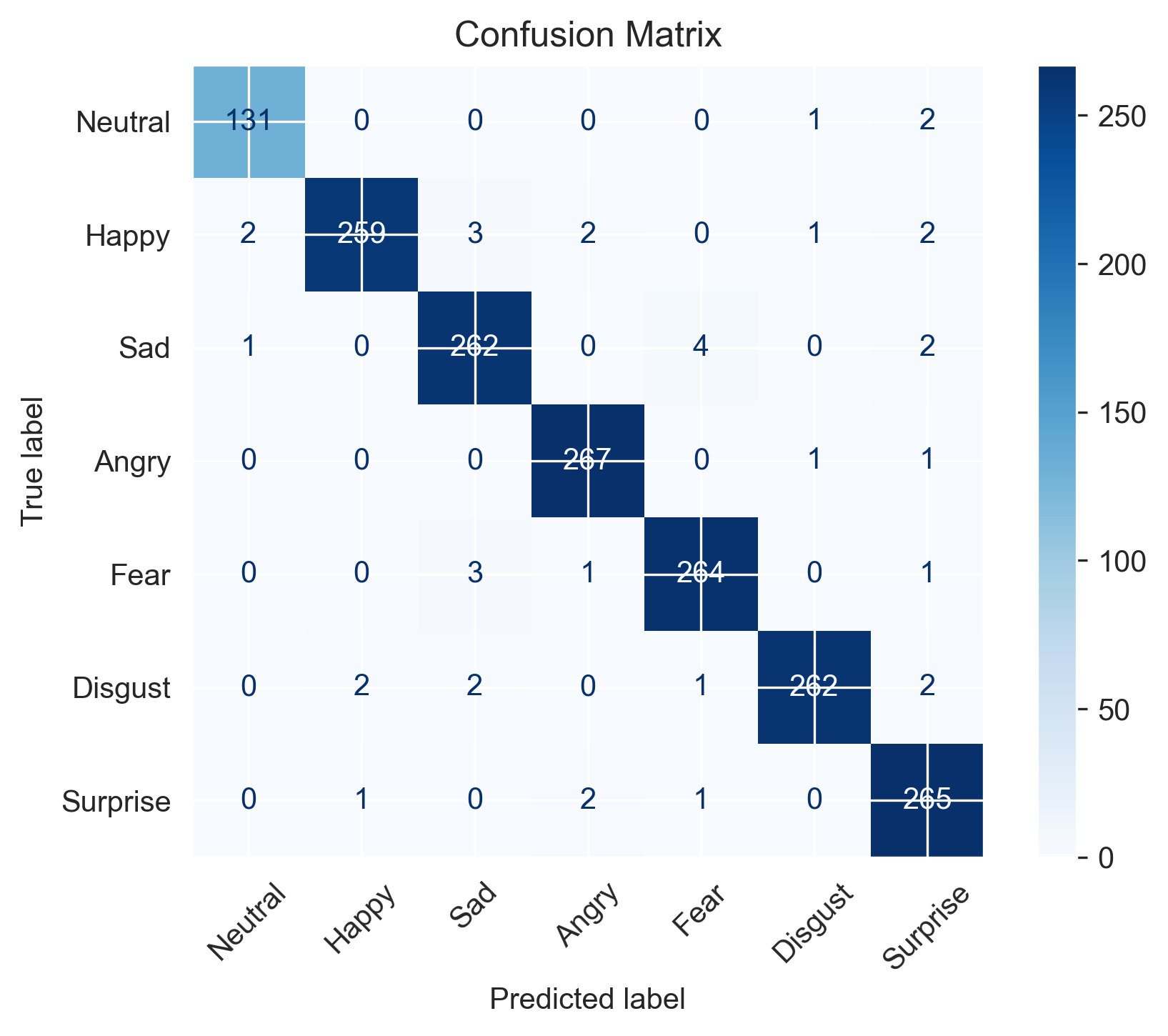}\hfill
  \includegraphics[width=0.25\linewidth]{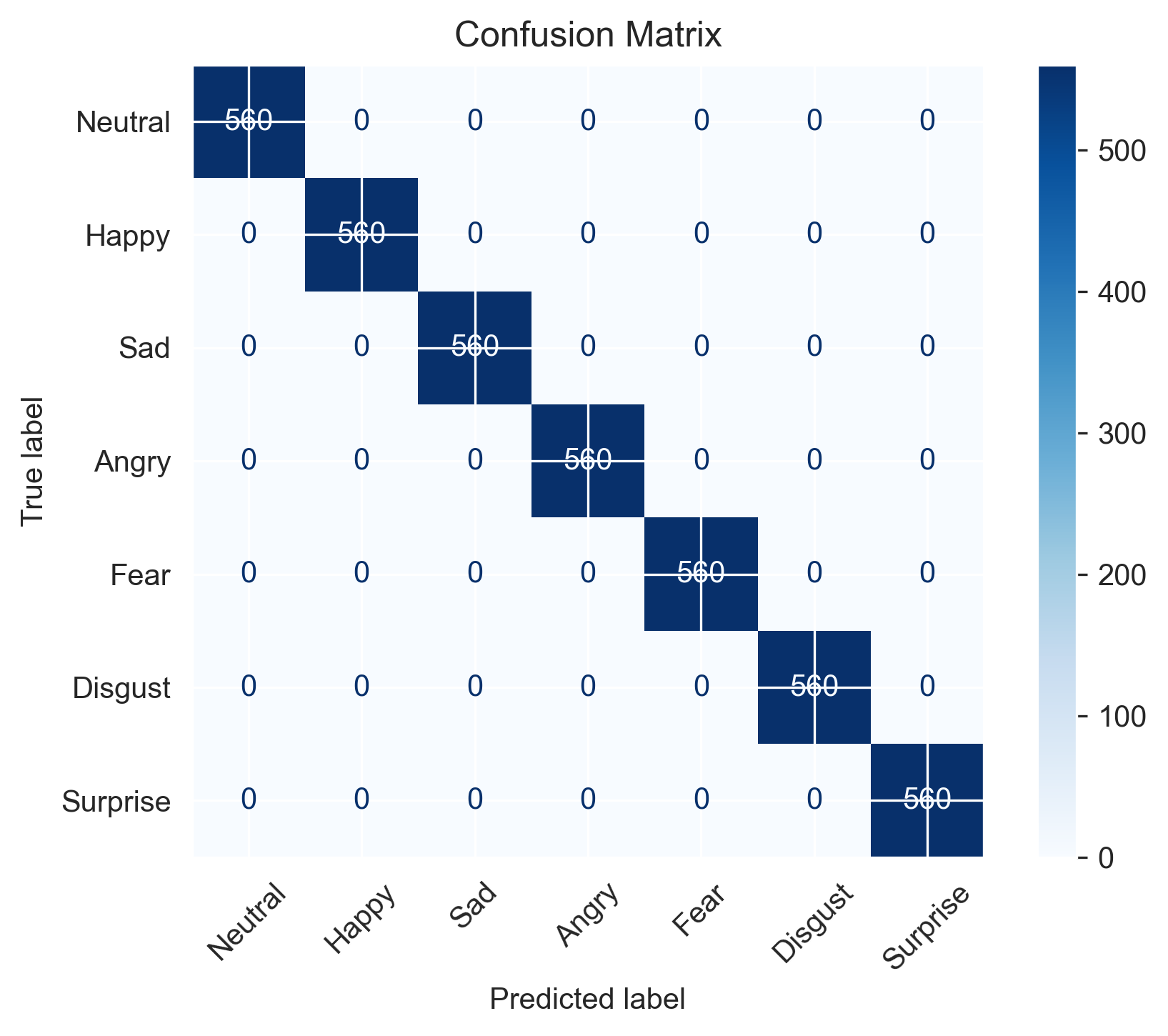}\hfill
  \includegraphics[width=0.25\linewidth]{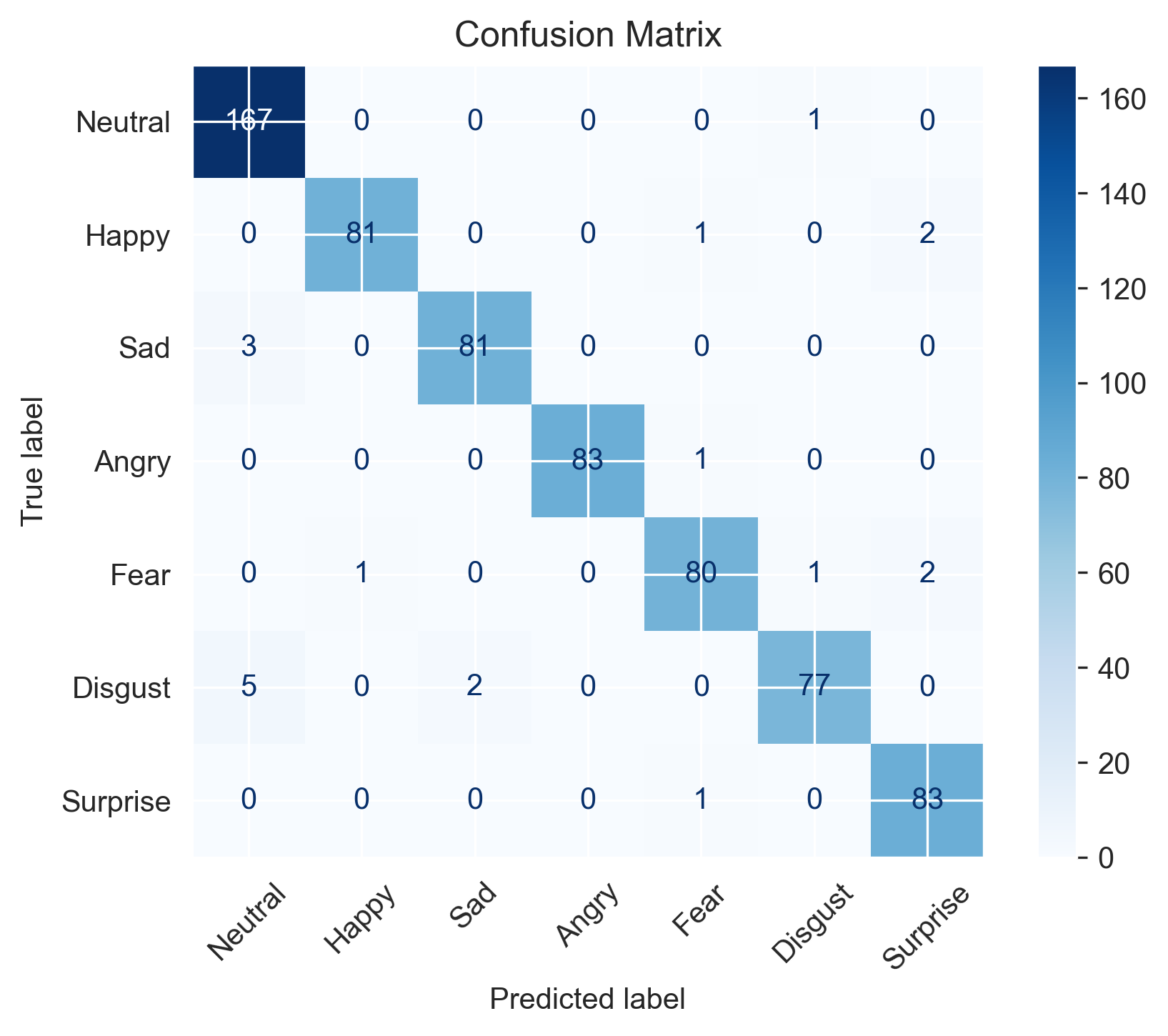}\hfill
  \includegraphics[width=0.25\linewidth]{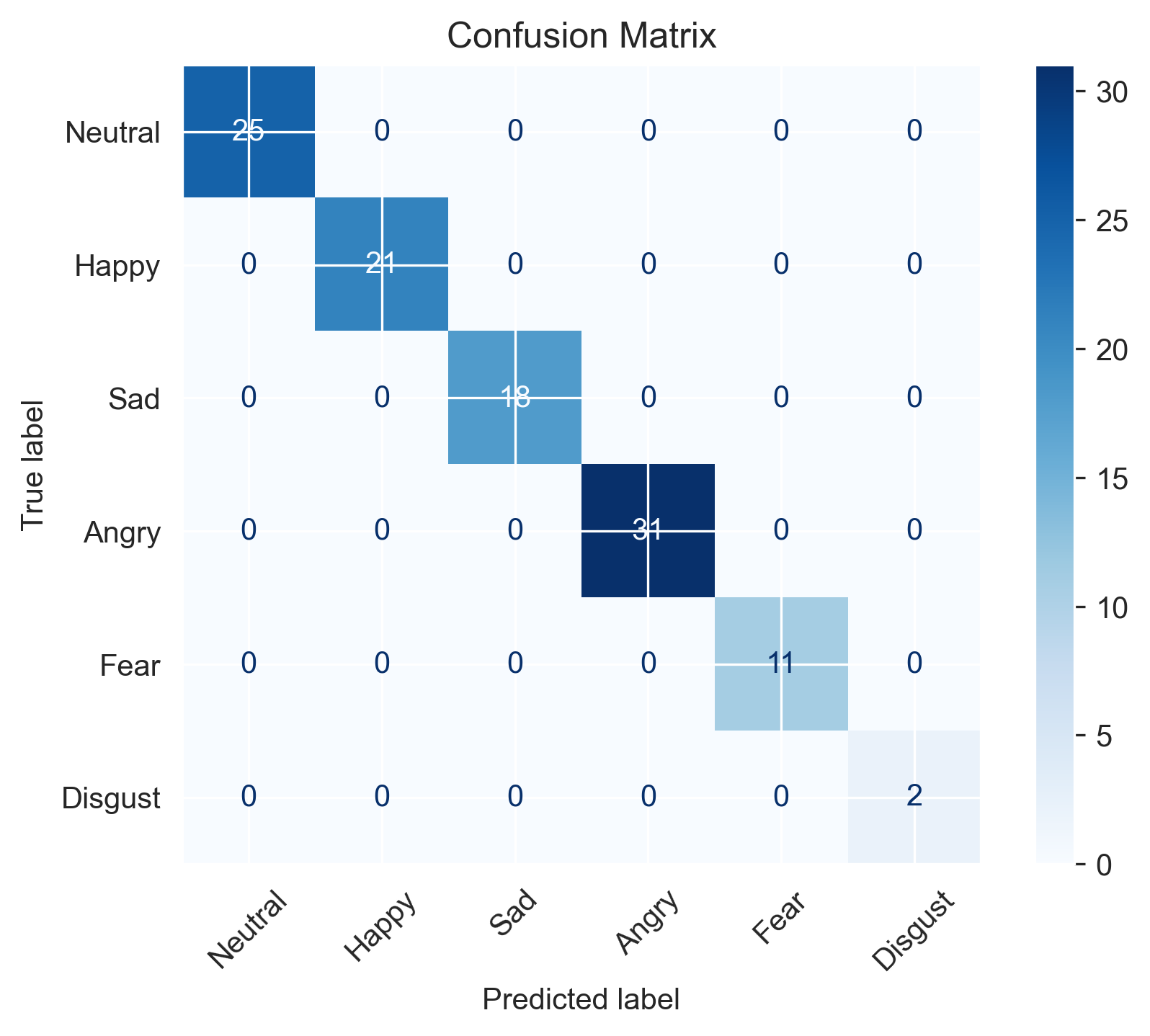}

  %\par\vspace{1em}  % Add space between the rows

  % --------- Row 2: 3 Images ----------
  \includegraphics[width=0.25\linewidth]{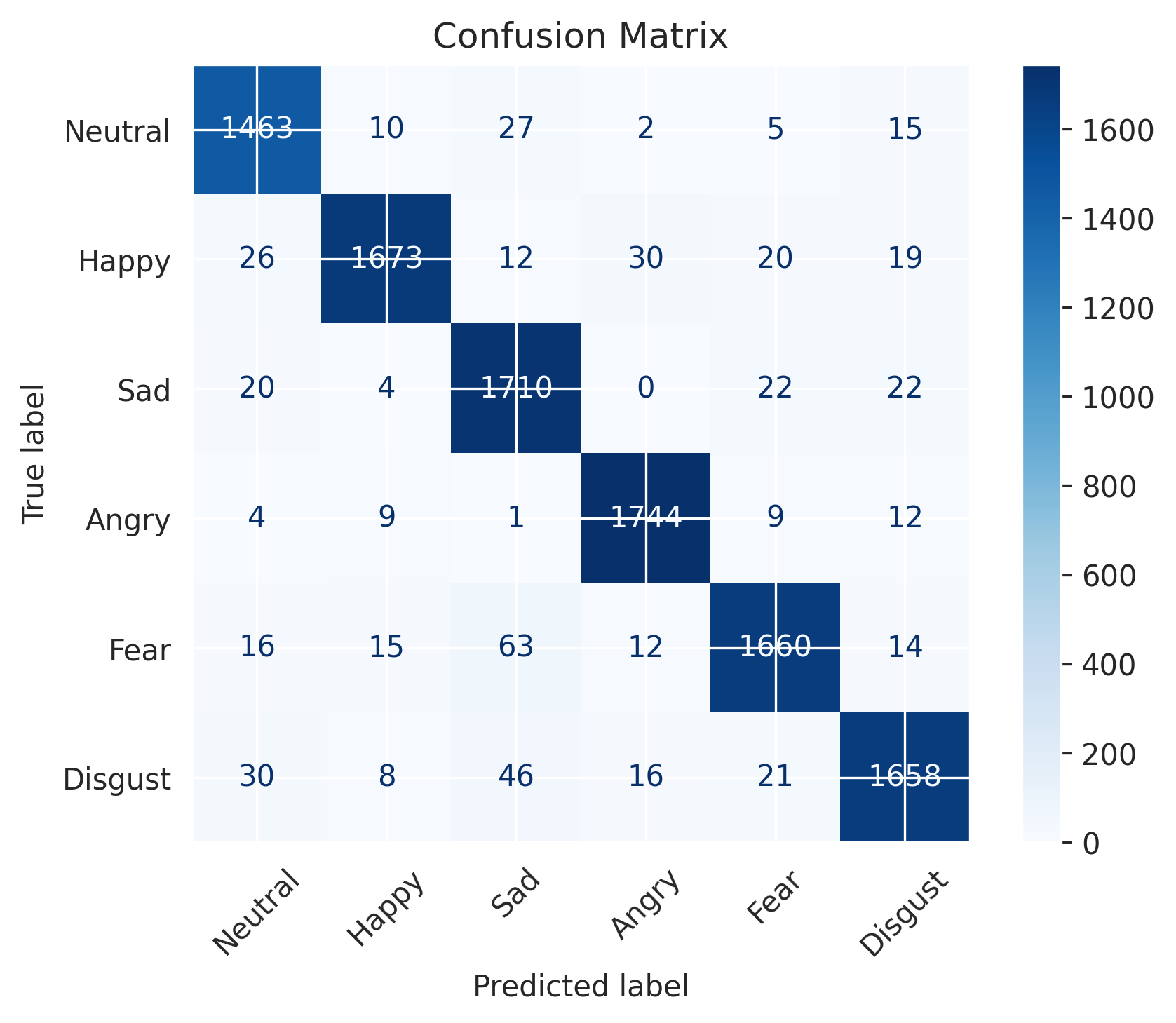}\hspace{2em}
  \includegraphics[width=0.25\linewidth]{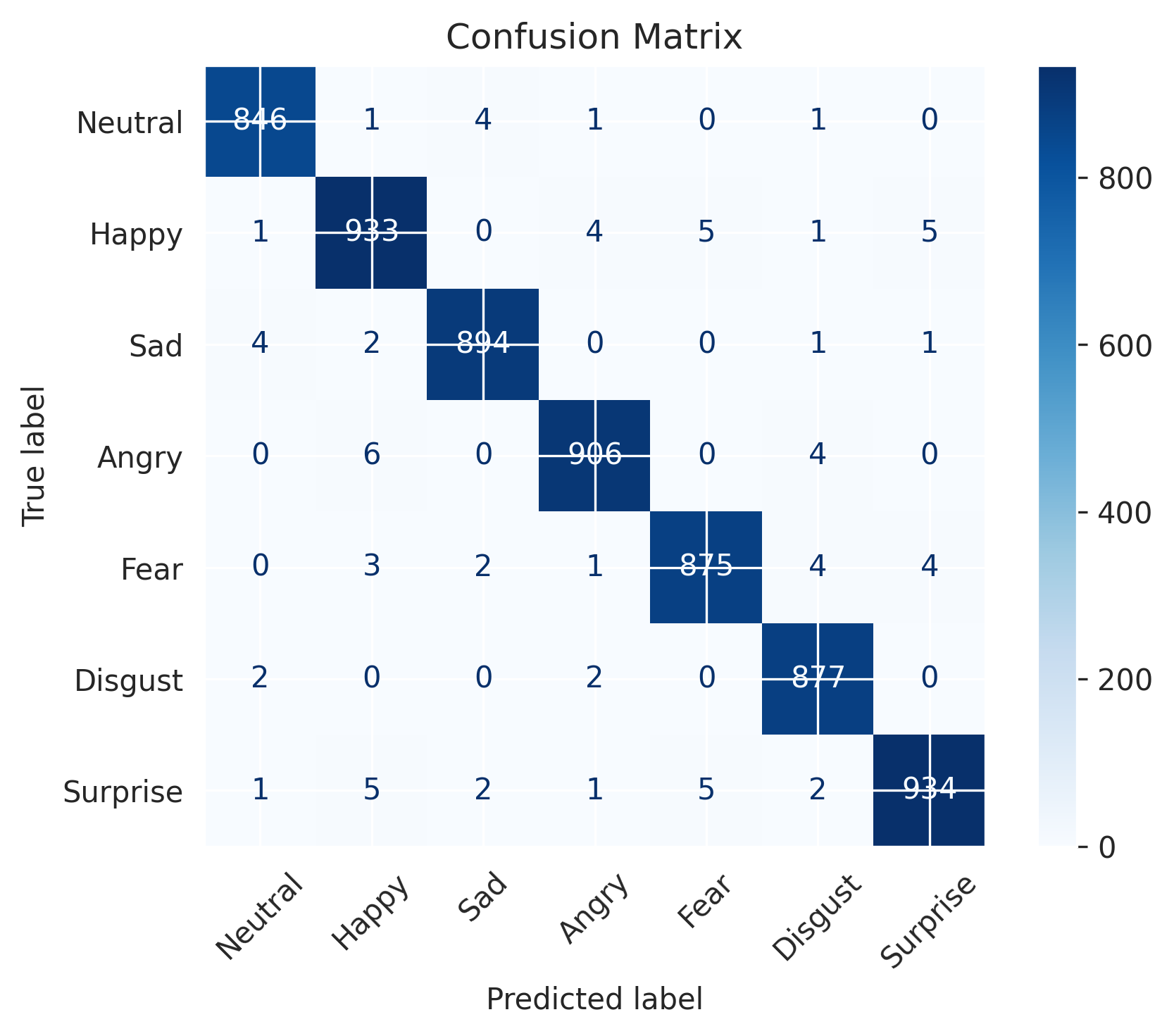}\hspace{2em}
  \includegraphics[width=0.25\linewidth]{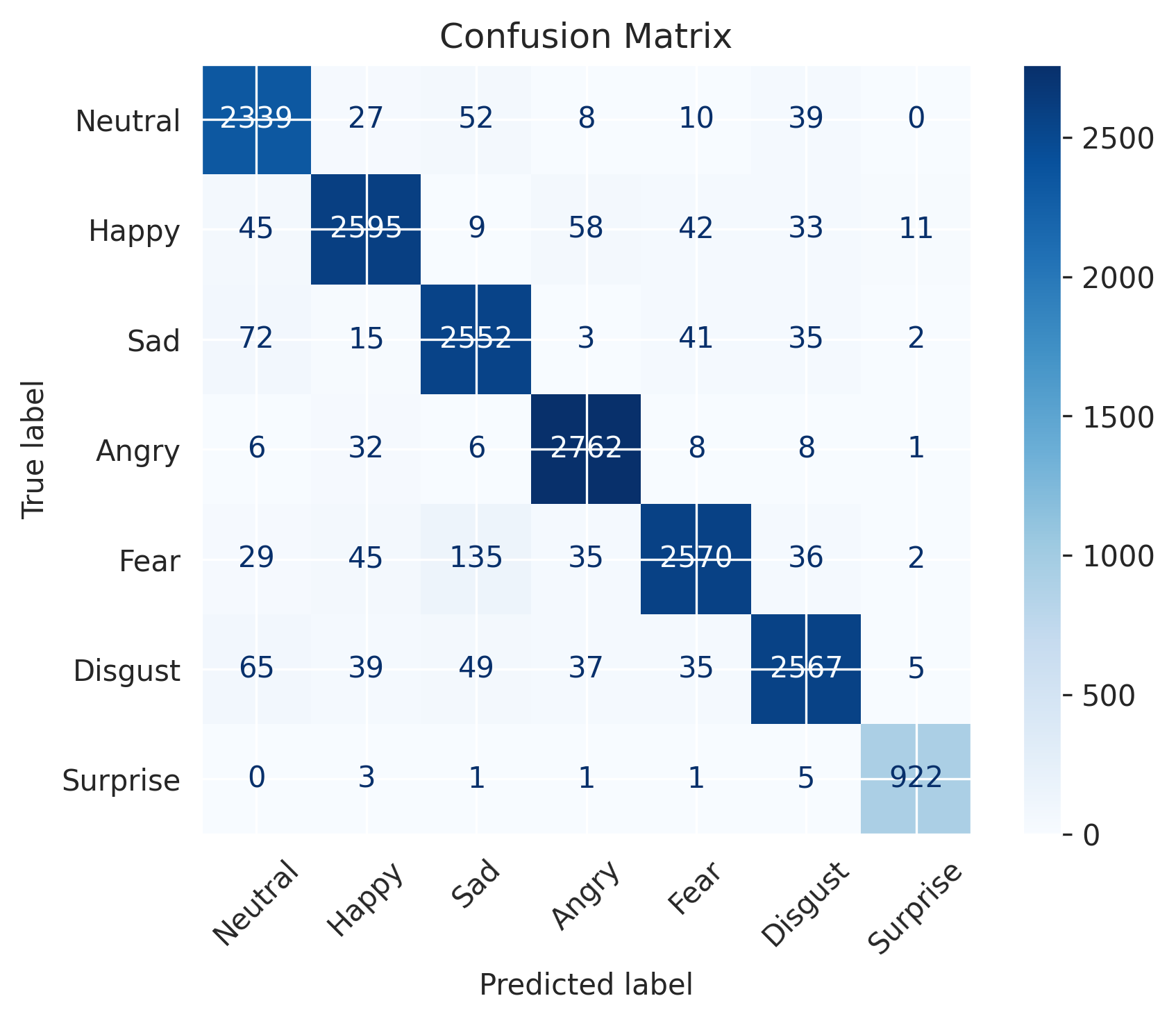}

  \caption{Confusion matrices of the DCRF-BiLSTM model on individual datasets (RAVDESS, TESS, SAVEE, EmoDB, CREMA-D) and combined datasets (R+T+S, R+T+S+E+C) top-to-bottom and left-to-right}
  \label{fig:cf_all}
\end{figure*}

A confusion matrix shows the number of correct and incorrect predictions for each class. The confusion matrix for all the emotions (classes) is given in 
%\textbf{\textit{\textcolor{blue}{
Figure~\ref{fig:cf_all}
%}}}. 
The emotion ‘surprise’ has the highest no. of correctly predicted samples for Ravdess, ‘disgust’ has the highest no. of correctly predicted samples for Tess, ‘neutral’ has the highest no. of correctly predicted samples for Savee, ‘angry’ has the highest no. of correctly predicted samples for EmoDB and Crema-D.

\vspace{-10pt} % Reduces space before the paragraph

% confusion Matrix images....................................

\subsection{Comparison with the State of the Art}
%\textbf{\textit{\textcolor{blue}{
Table~\ref{tab:comparison_accuracy}
%}}}
presents a comparison between the proposed DCRF-BiLSTM model and other related studies using the RAVDESS, TESS, SAVEE, EMO-DB, CREMA-D, and combined datasets. The evaluation is based on the average accuracy achieved in different experiments. The proposed model achieves the highest accuracy across all datasets and significantly outperforms previous methods. In particular, for the combined RAVDESS, TESS, and SAVEE (R+T+S) dataset, the model achieves an accuracy of 98.82\%, and for the combined RAVDESS, TESS, SAVEE, EMO-DB, and CREMA-D (R+T+S+E+C) dataset, the model achieves an accuracy of 93.76\%. Using an 80\%-20\% train-test split, the model demonstrates superior performance compared to existing approaches. The comparison is based on the average classification accuracy achieved by each model using various architectures and experimental setups.

The proposed DCRF-BiLSTM model consistently outperforms other methods across most of the datasets. It achieves the highest accuracy on RAVDESS (97.83\%), TESS (100.00\%), EMO-DB (100.00\%), and SAVEE (97.02\%) and the combined R+T+S dataset (98.82\%), surpassing previous benchmarks. %Although \cite{ahmed2023ensemble} achieved slightly higher performance on Crema-D (95.42\%) and the combined dataset (98.94\%), the proposed model remains highly competitive while offering robustness across all datasets.
%Added by Bithi
Furthermore, the proposed DCRF-BiLSTM model achieves an accuracy of 99.18\% on the combined RAVDESS, TESS, and SAVEE (R+T+S) datasets, and 94.03\% on the combined RAVDESS, TESS, SAVEE, EMO-DB, and CREMA-D (R+T+S+E+C) datasets using 5-fold cross-validation, further confirming its robustness and generalization across diverse emotional speech corpora.

Although \cite{ben2024enhancing} achieved slightly higher performance on the Savee dataset (97.11\%) and achieved the same performance on the combined dataset (99.18\%), our proposed model remains highly competitive while demonstrating robustness across all datasets.
Furthermore, there is no other model found in literature that worked with the combined five datasets, where our model achieved an accuracy of 93.76\%.

Moreover, in comparison to ensemble-based architectures such as the 1D CNNs-LSTM-GRU model by \cite{ahmed2023ensemble}, and ConvLSTM models \cite{ben2024enhancing}, the proposed DCRF-BiLSTM architecture—combining deep contextualized BiLSTM layers with a CRF-based sequence labeling mechanism—demonstrates superior generalization for audio signals. This confirms the model’s ability to capture temporal dependencies and contextual emotional insights more effectively. The proposed model shows  consistent performance for all datasets, ensuring the robustness and adaptability of different emotional speech corpora.

\begin{table*}[h!]
\centering
\small
\renewcommand{\arraystretch}{1.2}
\caption{Comparison of accuracy values (\%) of the proposed DCRF-BiLSTM approach with state-of-the-art work. Bold values indicate the best result of accuracy.}
\label{tab:comparison_accuracy}

\resizebox{\textwidth}{!}{%
\begin{tabular}{
>{\raggedright\arraybackslash}p{3.5cm} | 
c | c | c | c | c | c | c | 
>{\raggedright\arraybackslash}p{5.5cm}
}
\hline
\textbf{Paper} & \textbf{RAVDESS} & \textbf{TESS} & \textbf{SAVEE} & \textbf{EmoDB} & \textbf{CREMA-D} & \textbf{R+T+S} & \textbf{R+T+S+E+C} & \textbf{Model Info} \\
\hline
S. Mekruksavanich\cite{mekruksavanich2020negative}  &  75.83  & 55.71  & 65.83  &  -  &  65.77 & - & - & DCNN \\
\hline
U. Asiya \citep{asiya2021speech}  & 68.00 & - & - & - & - & 89.00  & - & 1D CNN \\
\hline
D. Issa \cite{issa2020speech}	  & 71.61  & -	    & -      &	86.1	& -	 & - 	  & - &	1D CNN LSTM, 5-fold cross validation \\ \hline
L. T. C. Ottoni \cite{ottoni2023deep}	  & 97.01  & \textbf{100.00}  & 90.62 &	 - 	& 83.28 & 97.37 & - &		2 CNN+LSTM, Split (80-10-10)\\ \hline
S. Jothiman \citep{jothimani2022mff}  &  92.60  &  99.60 &  84.90  &  -   &  89.90 &  - & - & CNN + LSTM model using the 80\%-20\% train-test split \\
\hline
M. R. Ahmed \citep{ahmed2023ensemble} & 95.62 & 99.46 & 93.22 & 95.42 & 74.09 & - & - & 1D CNNs-LSTM-GRU-ENSEMBLE \\
\hline
R. M. Ben-Sauo \citep{ben2024enhancing}  &  96.83  & 99.99  & \textbf{97.11} & 98.65  &  - & \textbf{99.18} & - & CONVLSTM model using the 5-fold-cross validation \\

\hline
\multirow{2}{*}{\parbox{3.5cm}{\centering \textbf{Proposed (DCRF-BiLSTM)}}} 
& 97.83 & \textbf{100.00} & 97.02 & \textbf{100.00} & \textbf{95.10} & 98.82 & 93.76 & Stacked BiLSTM + DeepCRF model using the 80\%-20\% train-test split \\
\cline{2-9} % Draw line only from 2nd to last column
& \textbf{97.99} & 99.99 & 96.10 & 99.26 & 94.41 & \textbf{99.18} & \textbf{94.03} & Stacked BiLSTM + DeepCRF model using 5-fold-cross validation \\

\hline
\end{tabular}
}
\end{table*}

\vspace{-20pt} % Reduces space before the paragraph
\section{Conclusions}
This study utilized a data preprocessing technique and three types of data augmentation techniques to expand the training and test datasets. We combined five diverse datasets to enhance generalizability and reduce potential bias from any single corpus. Still, some cross-dataset variation remains, which we aim to address in future work through domain adaptation techniques. Our SER also represented different kinds of important audio features, MFCC, Chroma, LMS, Spectral Contrast, RMSE, and ZCR, with a total of 190 features to reduce the complexity of the data and capture the most relevant and important meaningful information. Our proposed DCRF-BiLSTM hybrid framework, which uses the DeepCRF combined with the stacked Bidirectional LSTM, effectively captures the spatial and temporal information of audio signals. Evaluation of our model was performed on five publicly available datasets: RAVDESS, TESS, SAVEE, EMO-DB, and CremaD, as well as a combination dataset of (R+T+S) and (R+T+S+E+C). It shows the highest accuracy of 100.00\% on TESS and EmoDB , accurately recognizing and classifying speech emotions.

Although the proposed DCRF-BiLSTM architecture performed well, there are still areas for future work. Evaluating the model across different languages could provide insight into its adaptability and robustness. Incorporating contextual information, such as speaker identity, along with attention-based mechanisms, may help capture more complex emotional patterns. Since Bi-LSTM relies on future context, it is not well-suited for real-time audio sequences. Therefore, exploring alternative models that are more efficient for real-time speech processing would be beneficial. In future work, we plan to incorporate structured sentiment knowledge, such as SenticNet or emotion ontologies, to enhance reasoning over nuanced emotional expressions. We also aim to extend this framework to ASD-related emotion recognition and the early detection of emotional cues in children with ASD.

\section*{Acknowledgment}
This research is supported in part by the Public University Partnership Program at the Louisiana Department of Health, Bureau of Health Services Financing, through the National Institutes of Health (NIH) Award No. 10000411 and the Data Science Block Grant from the National Institute of General Medical Sciences, NIH Grant No. 2P20GM103424-24.

\textbf{Disclaimer.} This content is the sole responsibility of the authors and does not necessarily represent the official views of the Louisiana Department of Health.

\section*{Declaration on Generative AI}
The machine learning code in this research was implemented by the author(s), with limited assistance from ChatGPT for error checking and improvement. In addition, AI tools such as ChatGPT and Grammarly were used for grammar and spelling checks, paraphrasing, and rewording during the preparation of this manuscript. After using these tools, the author(s) reviewed and edited all content as needed and assume(s) full responsibility for the content of the publication.

\nocite{*}

\bibliography{references}

%\section{Appendix}
%\appendix{Appendix-A}
\appendix{Appendix-A}
\section{Appendix}

\begin{figure}[ht]
  \centering
  \includegraphics[width=1.0\linewidth]{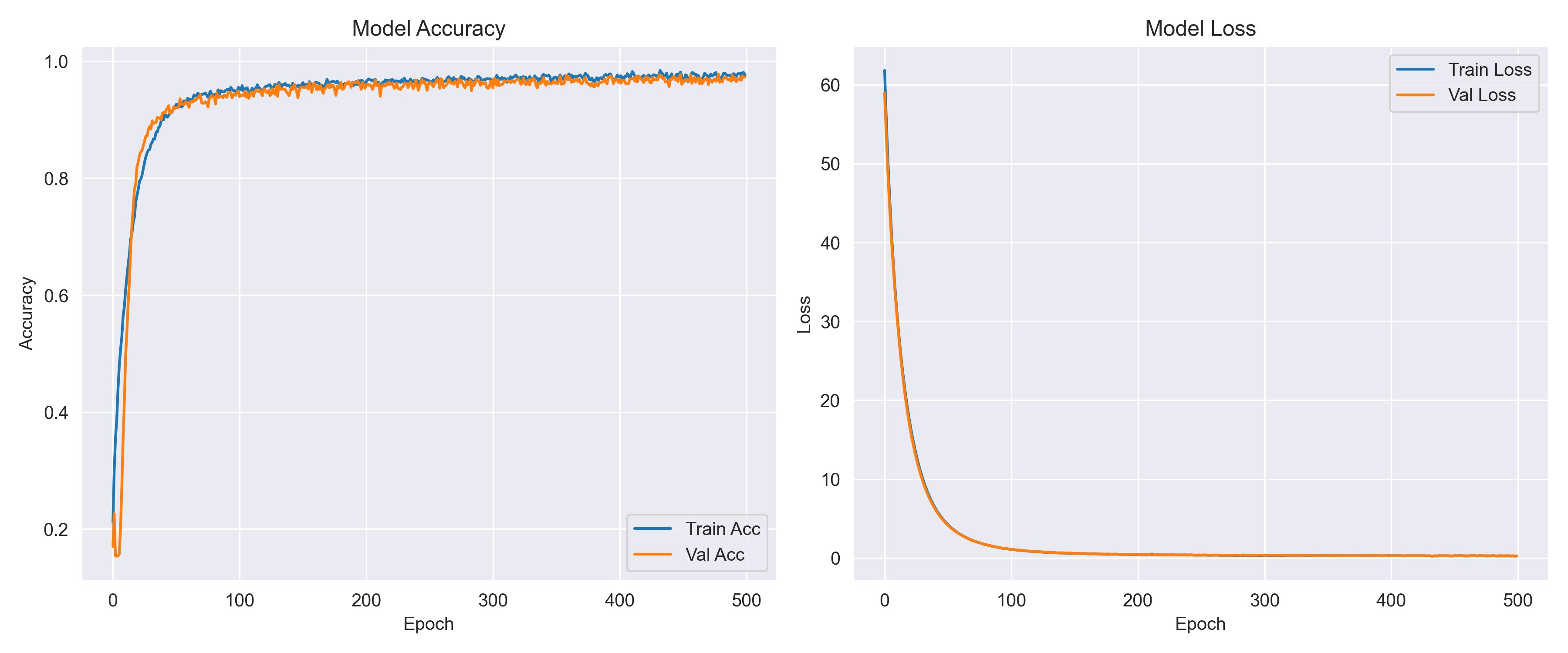}
  \caption{Training vs. validation accuracy and loss for the RAVDESS dataset.}
  \label{fig:ravdess_curve}
\end{figure}

\begin{figure}[ht]
  \centering
  \includegraphics[width=1.0\linewidth]{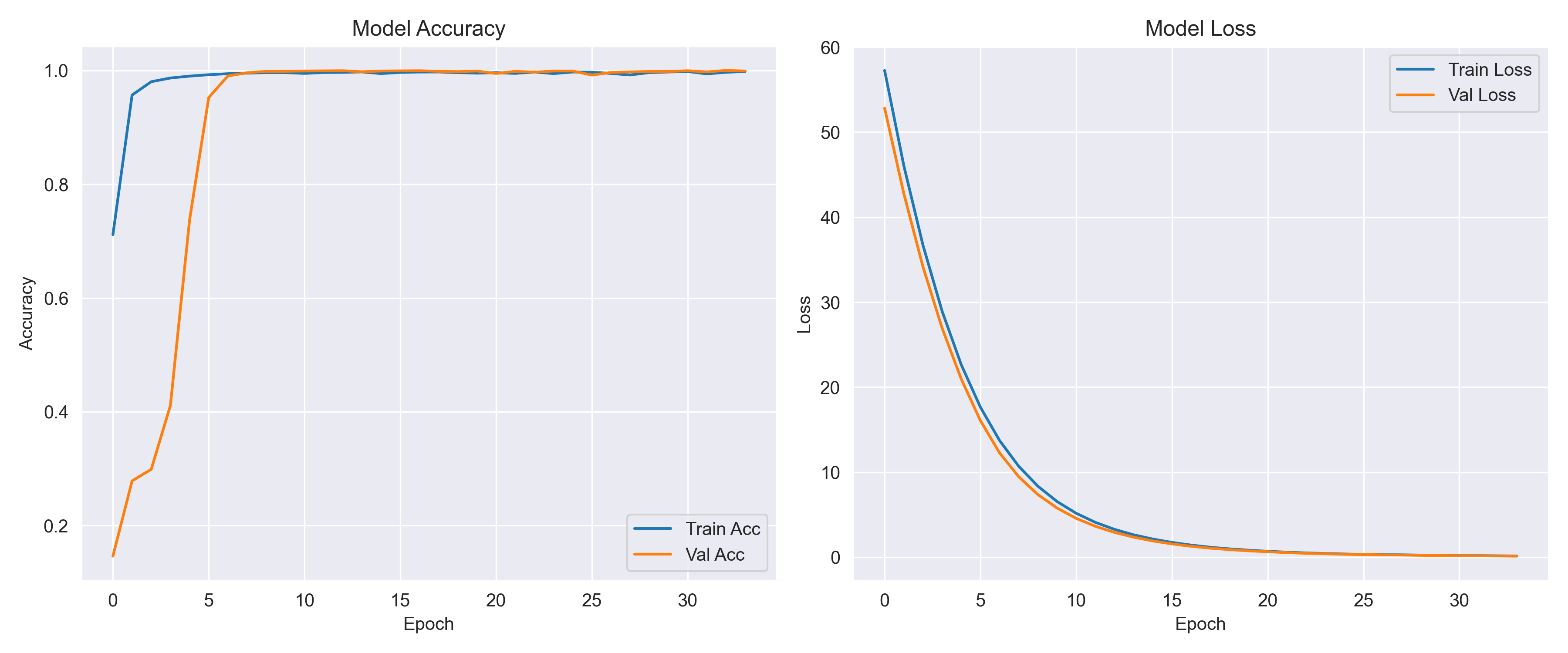}
  \caption{Training vs. validation accuracy and loss for the TESS dataset.}
  \label{fig:tess_curve}
\end{figure}

\begin{figure}[ht]
  \centering
  \includegraphics[width=1.0\linewidth]{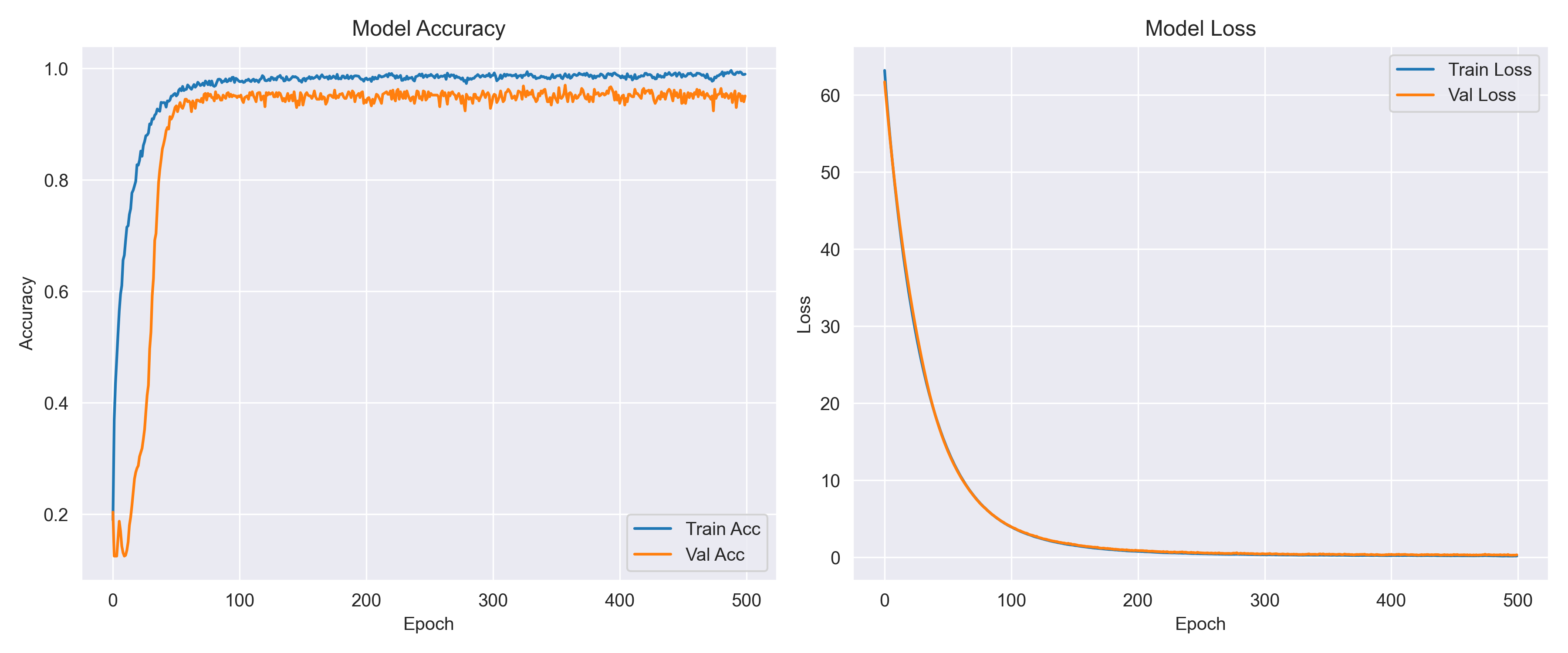}
  \caption{Training vs. validation accuracy and loss for the SAVEE dataset.}
  \label{fig:savee_curve}
\end{figure}

\begin{figure}[ht]
  \centering
  \includegraphics[width=1.0\linewidth]{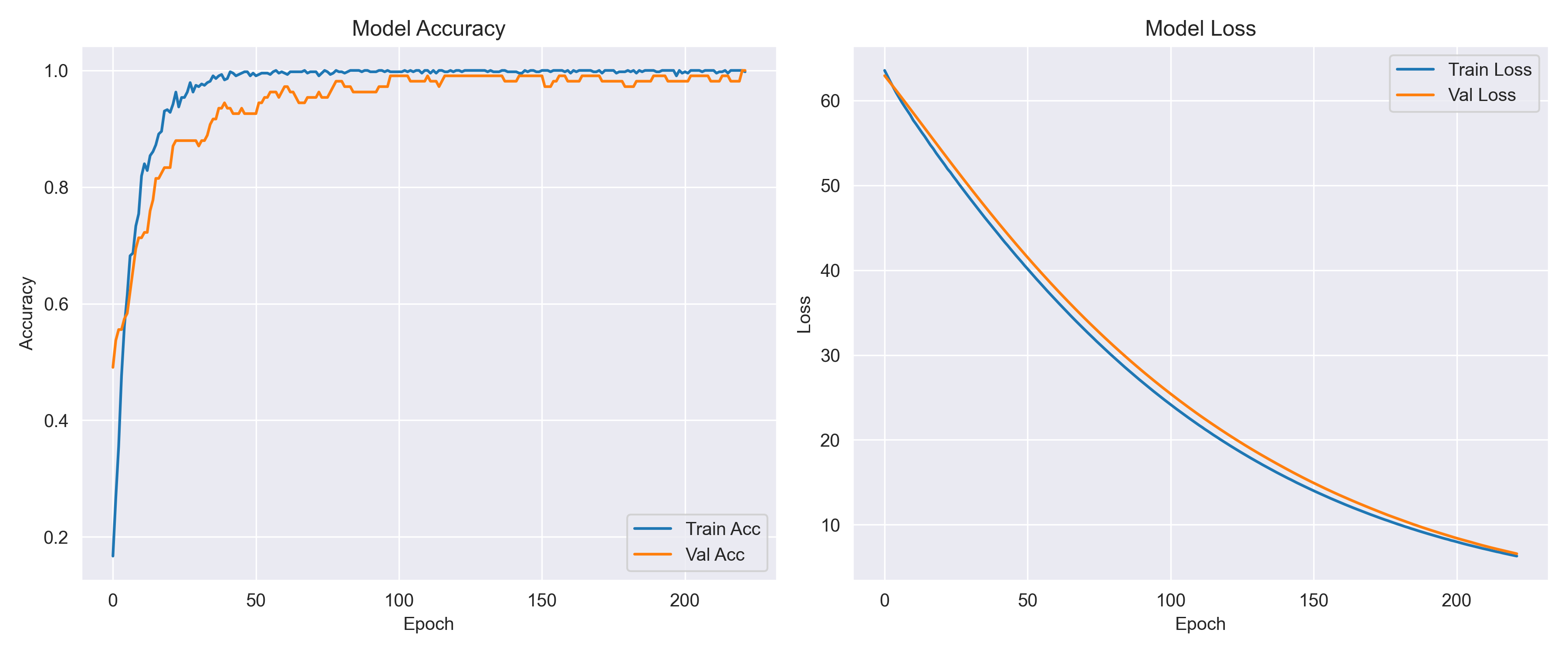}
  \caption{Training vs. validation accuracy and loss for the EmoDB dataset.}
  \label{fig:emodb_curve}
\end{figure}

\begin{figure}[ht]
  \centering
  \includegraphics[width=1.0\linewidth]{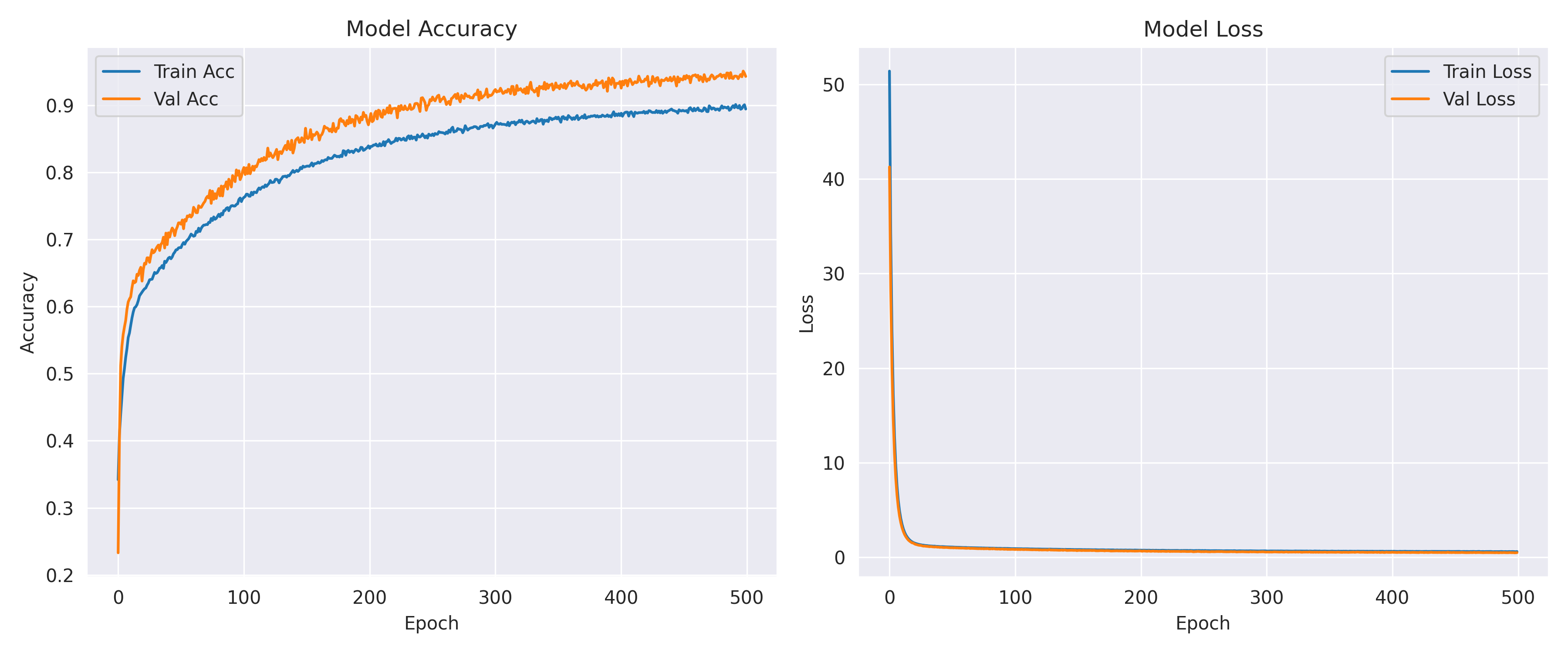}
  \caption{Training vs. validation accuracy and loss for the CREMA-D dataset.}
  \label{fig:cremad_curve}
\end{figure}

\end{document}